\documentclass[12pt,a4paper]{article}%
\pdfoutput=1
\usepackage{jheppub}
\usepackage{amsmath,amssymb,amsfonts,wasysym}
\usepackage{mathrsfs}
\usepackage[bf,footnotesize]{caption2}
\usepackage{bbm}
\usepackage[]{hyperref}%
\hypersetup{
colorlinks=true,
linkcolor=red,
anchorcolor=black,
citecolor=blue,
filecolor=cyan,
menucolor=red,
runcolor=filecolor,
urlcolor=magenta,
bookmarksnumbered=true
}
\allowdisplaybreaks
\def\l{\label}
\def\La{\mathcal{L}}

\def\({\left(}
\def\){\right)}
\def\f{\frac}
\def\be{\begin{equation}}
\def\ee{\end{equation}}
\def\bry{\begin{array}}
\def\ery{\end{array}}
\def\bes{\begin{subequations}}
\def\ees{\end{subequations}}
\def\bit{\begin{itemize}}
\def\eit{\end{itemize}}
\def\ben{\begin{enumerate}}
\def\een{\end{enumerate}}
\def\dst{\displaystyle}
\def\s{\sigma}

\def\g{\gamma}
\def\G{\Gamma}

\newcommand{\Dsl}{D\llap{/\kern+1.5pt}}
\newcommand{\MET}{E\llap{/\kern1.5pt}_T}

\def\nn{\nonumber}

\title{Higgs decay with monophoton$\,+\,\MET\,$signature\\
from low scale supersymmetry breaking}
\author[a]{Christoffer Petersson,}
\author[a,b]{Alberto Romagnoni,}
\author[a]{Riccardo Torre}
\affiliation[a]{Instituto de F\'isica Te\'orica UAM/CSIC}
\affiliation[b]{Departamento de F\'isica Te\'orica UAM \vspace{0.2cm}}
\affiliation{Universidad Aut\'onoma de Madrid, Cantoblanco, E-28049, Spain \vspace{0.5cm}}
\emailAdd{christoffer.petersson@csic.es}
\emailAdd{alberto.romagnoni@uam.es}
\emailAdd{riccardo.torre@csic.es}
\proceeding{\vspace{-1.65cm}\flushright{\small IFT-UAM/CSIC-12-23\\
FTUAM-12-86\\
LPN12-041\\}}
\abstract{We study the decay of a standard model-like Higgs boson into a gravitino and a neutralino, which subsequently decays promptly into another gravitino and a photon.  Such a decay can be important in scenarios where the supersymmetry breaking scale is of the order of a few TeV, and in the region of low transverse momenta of the photon, it may provide the dominant contribution to the final state with a photon and two gravitinos. We estimate the relevant standard model backgrounds and the prospects for discovering this Higgs decay through a  photon and missing transverse energy signal at the LHC in terms of a simplified model. By promoting the standard MSSM soft terms to supersymmetric operators, involving a dynamical goldstino supermultiplet, the parameters of the simplified model are related to the MSSM parameters and  the estimated discovery limits are used in order to constrain the parameter space. We show that it is possible to accommodate a SM-like CP-even neutral Higgs particle with a mass of 125 GeV, without requiring substantial radiative corrections, and with couplings sufficiently large for a signal discovery through the above mentioned Higgs decay channel with the upcoming data from the LHC. }
\begin{document} 
\maketitle
\setcounter{page}{2}
\section{Introduction}

Supersymmetric extensions of the Standard Model (SM) have the potential to stabilize the ElectroWeak (EW) scale and dynamically explain why it is hierarchically smaller than the Planck scale. 
However, the lack of experimental signals for supersymmetry (SUSY), as well as a conclusive argument that selects a particular scale or mechanism for SUSY breaking, motivates a broadening of the class of models which is being investigated. In this paper we study the less conventional scenario where the SUSY breaking scale $\sqrt{f}$ is of the order of a few TeV.  

A model-independent consequence of the spontaneous breaking of (global) SUSY is the existence of a Goldstone fermion, the goldstino. In the presence of gravity, SUSY is a local symmetry, the spin $1/2$ goldstino is eaten by the spin $3/2$ gravitino, becoming its longitudinal component and the gravitino acquires a mass $m_{3/2}= f/(\sqrt{3}M_{\mathrm{P}})$, where $M_{\mathrm{P}}=2.4\cdot 10^{18}$ GeV is the Planck mass. When $\sqrt{f}$ is of the order of a few TeV the gravitino is approximately massless ($m_{3/2}\approx 10^{-3}$ eV) and, due to the supersymmetric equivalence theorem \cite{1977PhLB...70..461F,1988PhLB..215..313C}, it can be replaced by its goldstino components. Moreover, gravitational effects can be neglected and SUSY can be treated as an approximate global symmetry. 

We will consider the case where the goldstino $G$ is the Lightest SUSY Particle (LSP), R-parity is preserved and the lightest neutralino $\chi^0_1$ is the Next-to-LSP (NLSP), with a mass $m_\chi$ smaller than the mass $m_h$ of the lightest CP-even neutral Higgs particle. This implies that the only relevant decay mode of the NLSP neutralino is into a goldstino and a photon (or a $Z$ boson if kinematically allowed, i.e.~if $m_\chi >m_Z$). Since $\sqrt{f}$ is very low, such decays will always be prompt on collider time scales.\footnote{For discussions concerning promptly decaying neutralinos giving rise to final states involving two photons and missing transverse energy see, e.g., Refs.~\cite{Dimopoulos:1996ku,
Ambrosanio:1996gg,
Dimopoulos:1996yq,Baer:1996hx,
Shirai:2009kn,Meade:2009vc,Ruderman:2011vv,Kats:2011ul}.}  
For $m_\chi<m_h/2$, the Higgs particle can decay into two neutralinos, each of which subsequently decays promptly into a photon and a goldstino, giving rise to a final state consisting of two photons and missing transverse energy ($\MET$). 
This possibility, as well as the experimental bounds on light unstable neutralinos 
 (and charginos) have been discussed in Refs.~\cite{Mason:1207862,Mason:2011uy}.  We will restrict our analysis to the region where $m_h/2<m_\chi<m_h$.

The process we study in this paper arises from the decay of a SM-like Higgs boson $h$ into a gravitino and an NLSP neutralino, which subsequently decays promptly into another gravitino and a photon, $gg\to h\to \chi_{1}^{0}G\to\gamma GG$, giving rise to a final state consisting of a photon and $\MET$. Such a final state has been  discussed in several papers \cite{1982PhLB..117..460F,1985ZPhyC..27..577N,1986PhLB..175..471F,1991PhLB..258..231D,Lopez:1996ju,Lopez:1996fr,Mawatari:2011wl,Argurio:2011wl}, but  the  processes considered in those papers were $q\bar{q}/e^+ e^-\to \gamma GG$, involving the non-resonant exchange of a squark/selectron in the t-channel, and \mbox{$q\bar{q}/e^+ e^-\to Z\to\chi_{1} ^{0}G\to \gamma GG$}, involving the exchange of a $Z$ boson or a photon in the s-channel. In Refs.~\cite{1997hep.ph...11516B,Brignole:1998eg} such processes were considered in an effective theory where all superpartners, except for the goldstino, had been integrated out in order to set a model-independent lower bound on $\sqrt{f}$. The current experimental lower bound on $\sqrt{f}$ is around 300 GeV \cite{LCollaboration:2004hl,2002hep.ex....5057C}, but it is expected that the LHC will be able to probe $\sqrt{f}$ up to around 1.6 TeV \cite{Brignole:1998eg}. 

The t-channel process $q\bar{q}\to \gamma GG$ is relevant at the LHC in the region where the transverse momenta of the photon ($p_T^\gamma$) is high, of the order of 100 GeV or higher. Since the $p_T^\gamma$ distribution for the resonant Higgs process $gg\to h\to \chi_{1}^{0}G\to\gamma GG$ we are interested in have a kinematic end-point at $m_h/2$, we will only consider the region where $p_T^\gamma<m_h/2$, in which the t-channel process turns out to be subleading. In the region where $p_T^\gamma<m_Z/2$, also the resonant $Z$ boson process $q\bar{q}\to Z\to\chi_{1}^{0}G\to \gamma GG$ contributes. However, after the implementation of our kinematic cuts, chosen in order to optimize the signal significance, this contribution turns out to be negligible. The Higgs decay $h\to \chi_1^0 G$ was studied in Ref.~\cite{Djouadi:1997bx} (for $\sqrt{f}=650$ GeV, \mbox{$m_h=139$ GeV}), but the Branching Ratio (BR) for this decay was 
estimated to be too small in order to give rise to a significant signal. 
See also Ref.~\cite{Antoniadis:2010hs} for a discussion concerning the case when such a Higgs decay is invisible and Ref.~\cite{Gherghetta:2011tn} for when it takes place in scenarios where SUSY is broken at the TeV scale by strong dynamics.  

The paper is organized as follows. In order to study the $\gamma+\MET$ signal in a way which is as model-independent as possible, we begin in Section \ref{Sec:Pheno} by defining a simplified model that only contains the relevant particles, i.e.~the SM particles, the goldstino and the NLSP neutralino as well as the relevant interactions. 
We estimate the SM backgrounds for the $\gamma+\MET$ final state in the region $p_{T}^{\gamma}<m_{h}/2$ and we present prospects for discovering this Higgs decay mode at the LHC with \mbox{$\sqrt{s}=8$ TeV} and the expected integrated luminosity for the upcoming LHC run\footnote{It was recently decided to raise the center of mass energy from $7$ to $8$ TeV for the 2012 LHC run \cite{8TeV}. The expected total integrated luminosity is $5$ fb$^{-1}$ at $7$ TeV and around $15$ fb$^{-1}$ at $8$ TeV. For simplicity we will study the case of $20$ fb$^{-1}$ at $8$ TeV.}, i.e.~\mbox{$15\div 20$ fb$^{-1}$}. In particular we present these prospects by estimating the  sensitivity with the mentioned LHC parameters to the BR of $h\to \chi_{1}^{0}G$ as a function of the neutralino mass in the range $m_h/2<m_{\chi}<m_h$. Inspired by the recent LHC hints concerning the mass of the Higgs particle \cite{ATLASCollaboration:2012uh,CMSCollaboration:2012tb}, we will focus our analysis on the value \mbox{$m_h=125$ GeV}. 

In Section \ref{Sec:Model} we relate the parameters of the simplified model to those of the Minimal SUSY SM (MSSM) by considering a small set of higher dimensional supersymmetric operators involving the goldstino supermultiplet. Since these operators give rise both the usual MSSM soft terms and the goldstino interactions relevant for our simplified model, the coefficients of these interactions are determined by ratios of soft parameters over the SUSY breaking scale $f$. Concerning the couplings relevant for the Higgs decay under consideration, it is only a small set of supersymmetric operators that is relevant. This is in contrast to, for instance, the scalar and neutralino mass spectrum, which can be significantly affected by the presence of additional higher dimensional supersymmetric operators, see  Refs.~\cite{Brignole:2003hb,Dine:2007eg, Antoniadis:2007et, Antoniadis:2008cu, Carena:2009ey, Carena:2011ty, Boudjema:2011un, Boudjema:2012wp} for analysis, which are beyond the scope of this paper.
We show that it is possible  to raise the  tree level masses for all the Higgs scalar particles, for any value of $\tan\beta$, and also to  generate all the goldstino interactions relevant for the simplified model considered in Section \ref{Sec:Pheno} with the addition of the few operators giving rise to the MSSM soft terms. Moreover we expect that the addition of the complete set of higher dimensional operators can even improve the situation. In this scenario it is possible to accomodate a SM-like CP-even neutral Higgs particle with a mass of 125 GeV and with couplings that are large enough for a signal discovery through the process $gg\to h\to \chi_{1}^{0}G\to\gamma GG$ with the upcoming data from the LHC.

In Section \ref{Sec:Conclusion} we conclude with a summary of our results and discuss some future directions. In Appendix \ref{appendix}, the analytic formulae for the Higgs mass and the relevant couplings are provided.

\section{A simplified model for monophoton$\,+\,\MET\,$SUSY signals at the LHC}\l{Sec:Pheno}

In this section, we study the phenomenology of the photon$\,+\,\MET\,$ final state, arising from a low scale SUSY breaking scenario, by consider the following effective Lagrangian
\be\l{ph1}
\La_{\text{eff}}=\La_{\text{SM}}+\La_{\text{NP}}\,,
\ee
where $\La_{\text{SM}}$ is the usual SM Lagrangian and
\be\l{ph2}
\La_{\text{NP}} =  \f{m^{2}}{\sqrt{2}F}\Bigg[g_{h\chi} h\chi_{1}^{0}G+\f{g_{\chi \g}}{m}G\sigma^{\mu\nu}F_{\mu\nu} \chi_{1}^{0}+\f{g_{\chi Z\,1}}{m}G\sigma^{\mu\nu}Z_{\mu\nu}\chi_{1}^{0}+g_{\chi Z\,2}\bar{G}\bar{\s}^{\mu}Z_{\mu}\chi_{1}^{0}+\text{h.c.}\Bigg]
\ee
is the New Physics (NP) Lagrangian in Weyl notation. In Eq.~\eqref{ph2}, $F_{\mu\nu}$ and $Z_{\mu\nu}$ are the photon and the $Z$ boson field strengths, $h$ is the Higgs boson, $\chi_{1}^{0}$ is the NLSP neutralino  and $G$ is the goldstino. Also, the $g$'s are dimensionless couplings, \mbox{$\s^{\mu}=(\mathbbm{1}_{2},\s^{i})$}, $\bar{\s}^{\mu}=(\mathbbm{1}_{2},-\s^{i})$ and $\s^{\mu\nu}=i/4(\s^{\mu}\bar{\s}^{\nu}-\s^{\nu}\bar{\s}^{\mu})$, with $\s^{i}$ being the ordinary Pauli matrices. The two mass scales  $\sqrt{F}$ and $m$ can be interpreted as the SUSY breaking scale\footnote{We use the capital letter here in order to avoid confusion between the scale $\sqrt{F}$ used in the simplified model and the SUSY breaking scale $\sqrt{f}$ used in the rest of the paper.} and a scale related to the soft parameters, respectively. As can be seen from Eq.~\eqref{ph2}, the NP Lagrangian is only relevant when $\sqrt{F}$ is not too separated from $m$. Since $m$ is also related to the masses of the Higgs particle and the neutralino, $m_h$ and $m_\chi$, respectively, we expect $m$ to be of the order of the ElectroWeak (EW) scale. Thus, in order for the Lagrangian \eqref{ph2} to be phenomenologically relevant, $\sqrt{F}$ should be around the TeV scale.  

\begin{figure}[!t]
\begin{center}
\includegraphics[scale=0.36]{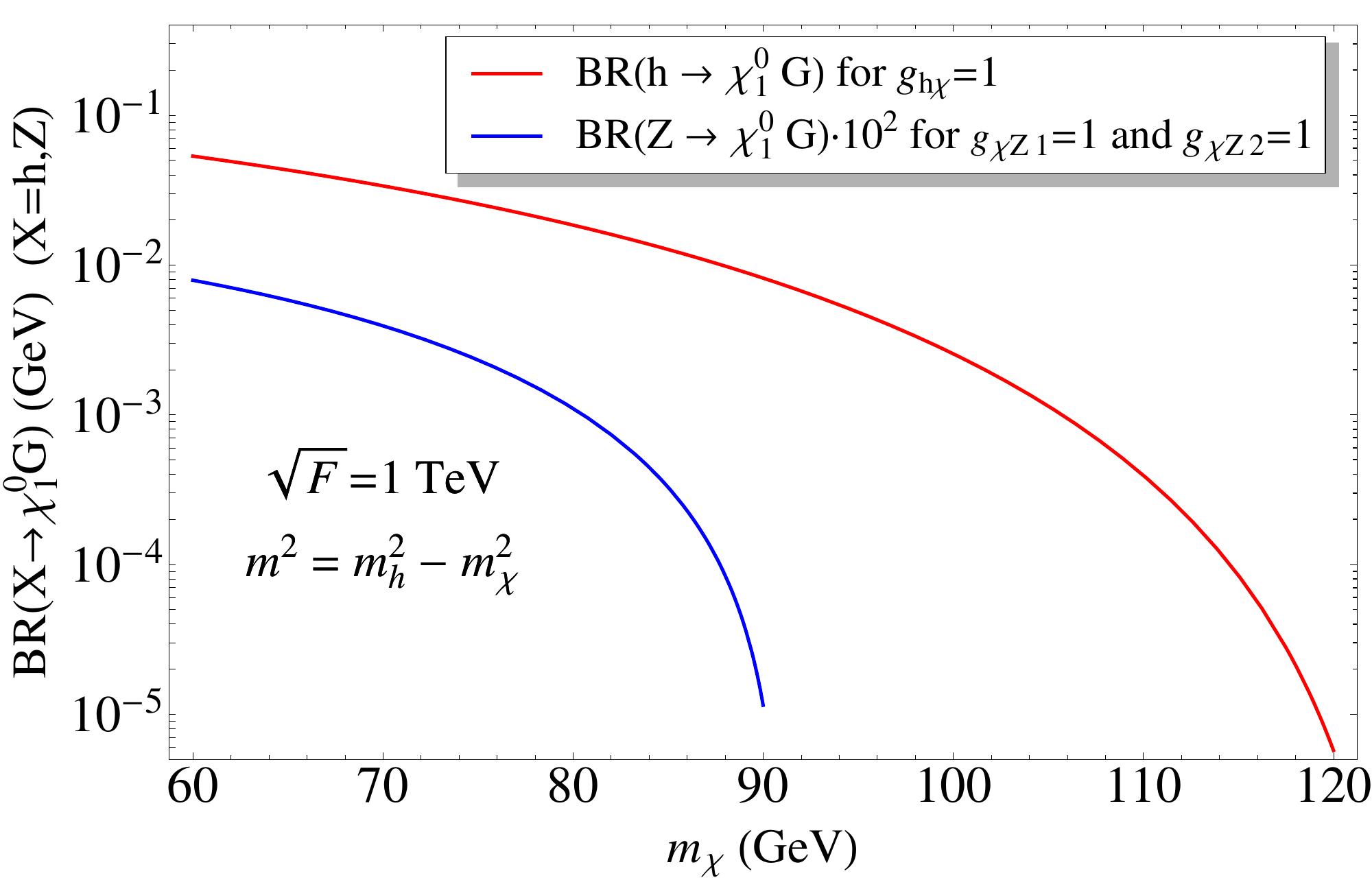}\hspace{2mm}
\includegraphics[scale=0.36]{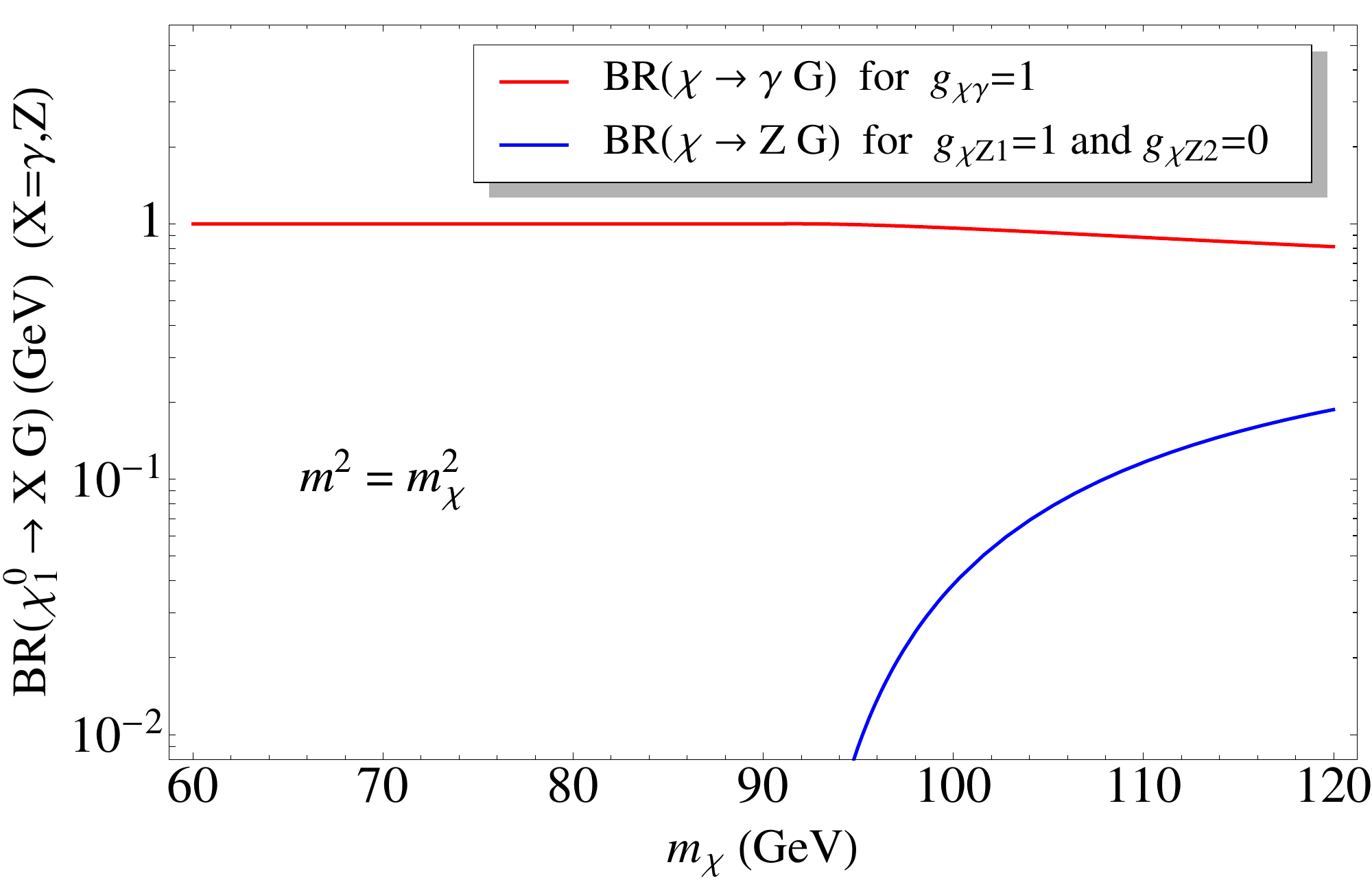}
\end{center}
\caption{ 
\small
Left panel: Higgs and $Z$ bosons decay widths into a neutralino $\chi_{1}^{0}$ and a goldstino $G$ as functions of the neutralino mass; right panel: BRs for the decay of the neutralino into a photon or a $Z$ boson and a goldstino as functions of the neutralino mass. We have set all the dimensionless couplings equal to one and the dimensionful parameters to be $m^{2}=m_{h}^{2}-m_{\chi}^{2}$ (see the text for details on this choice) and $\sqrt{F}=1$ TeV.
}\label{fig:widthBRs}
\end{figure}

When kinematically allowed, the Lagrangian \eqref{ph2} gives rise to new decay modes of the Higgs and the $Z$ bosons into a neutralino and a goldstino with the following partial widths,
\begin{eqnarray}
\Gamma(h\to \chi_{1}^{0}G) & = &  \f{m_{h}}{16\pi}\f{g_{h\chi}^{2}m^{4}}{F^{2}}\(1-\f{m_{\chi}^{2}}{m_{h}^{2}}\)^{2}\,,\l{ph3a} \\
\Gamma(Z\to \chi_{1}^{0}G) & = & \f{1}{24\pi m_{Z}}\(1-\f{m_{\chi}^{2}}{m_{Z}^{2}}\)\f{m^{4}}{F^{2}}\Bigg[g_{\chi Z\, 2}^{2}m_{Z}^{2}\(1-\f{m_{\chi}^{2}}{2m_{Z}^{2}}-\f{m_{\chi}^{4}}{2m_{Z}^{4}}\)\l{ph3b}\\
&& \dst +3g_{\chi Z\, 1}g_{\chi Z\, 2}\f{m_{\chi}m_{Z}^{2}}{m}\, \(1-\f{m_{\chi}^{2}}{m_{Z}^{2}}\)+g_{\chi Z\, 1}^{2}\f{m_{\chi}^{4}}{m^{2}}\(-1+\f{m_{Z}^{2}}{2m_{\chi}^{2}}+\f{m_{Z}^{4}}{2m_{\chi}^{4}}\)\Bigg]\nn \,. 
\end{eqnarray}
Moreover, Eq.~\eqref{ph2} provides the decay modes of the neutralino into a goldstino and a photon or, when kinematically allowed, a $Z$ boson, with the corresponding partial widths,
\begin{eqnarray}
\Gamma(\chi_{1}^{0}\to \g G) & = & \f{m_{\chi}^{3}}{16\pi}\f{g_{\chi \g}^{2}m^{2}}{F^{2}}\,,\l{ph3c}\\
\Gamma(\chi_{1}^{0}\to ZG) & = & \dst \f{1}{16\pi m_{\chi}}\(1-\f{m_{Z}^{2}}{m_{\chi}^{2}}\)\f{m^{4}}{F^{2}}\Bigg[g_{\chi Z\, 2}^{2}m_{Z}^{2}\(-1+\f{m_{\chi}^{2}}{2m_{Z}^{2}}+\f{m_{\chi}^{4}}{2m_{Z}^{4}}\) \l{ph3d} \\
&& \dst +3g_{\chi Z\, 1}g_{\chi Z\, 2}\f{m_{\chi}^{3}}{m}\, \(1-\f{m_{Z}^{2}}{m_{\chi}^{2}}\)+g_{\chi Z\, 1}^{2}\f{m_{\chi}^{4}}{m^{2}}\(1-\f{m_{Z}^{2}}{2m_{\chi}^{2}}-\f{m_{Z}^{4}}{2m_{\chi}^{4}}\)\Bigg]\nn \,. 
\end{eqnarray}
In order to have an idea of the order of magnitude of these partial widths we show in Fig.~\ref{fig:widthBRs} $\text{BR}(h\to\chi_{1}^{0}G)$ and $\text{BR}(Z\to\chi_{1}^{0}G)$ as well as $\text{BR}(\chi_{1}^{0}\to\g G)$ and $\text{BR}(\chi_{1}^{0}\to Z G)$ as functions of the neutralino mass in the range $60<m_{\chi}<120$ GeV with all the dimensionless couplings in Eq.~\eqref{ph2} set equal to one but $g_{\chi Z2}$ which is set equal to zero\footnote{The choice $g_{\chi Z2}=0$ corresponds to neglecting the coupling to the longitudinal component of the $Z$ boson and motivates the choice of the scale $m=m_{\chi}$. Also notice that the simple choice of setting all the other couplings to unity is only for illustrative purposes and generically not realized in explicit models.}. The value of the mass parameter $m$ has been chosen to be $m^{2}=m_{h}^{2}-m_{\chi}^{2}$ in the case of the Higgs decay and $m=m_{\chi}$ in the case of the neutralino decay. The former choice corresponds to the simple case of no mixing, where the lightest neutralino is purely Higgsino. The latter choice is motivated by the fact that $m_{\chi}$ is the only mass scale relevant for the decay of the neutralino into a photon and a goldstino and that the coupling to $F^{\mu\nu}$ and to $Z^{\mu\nu}$ should be the same up to weak rotations.

Here and in the following analysis we will set $m_{h}=125$ GeV, inspired by the recent LHC results of Refs.~\cite{ATLASCollaboration:2012uh,CMSCollaboration:2012tb}. 
Moreover, we will only consider the case where 
\be\l{ph4}
\Gamma(h\to \chi_{1}^{0}G)\ll \G^{h}_{\text{tot}}\,,\qquad \qquad \Gamma(Z\to \chi_{1}^{0}G)\ll \delta\Gamma_{Z}\,,
\ee
where $\G^{h}_{\text{tot}}$ is the total SM Higgs width and $\delta\Gamma_{Z}=2.3$ MeV is the present experimental uncertainty on the $Z$ boson width \cite{Nakamura:2010dr}. The relations \eqref{ph4} imply that the BRs of the Higgs and the $Z$ bosons into all the SM final states are not affected, within the experimental errors, by the new decay modes in Eqs.~\eqref{ph3a} and \eqref{ph3b}.

\begin{figure}[!t]
\begin{center}
\includegraphics[scale=0.45]{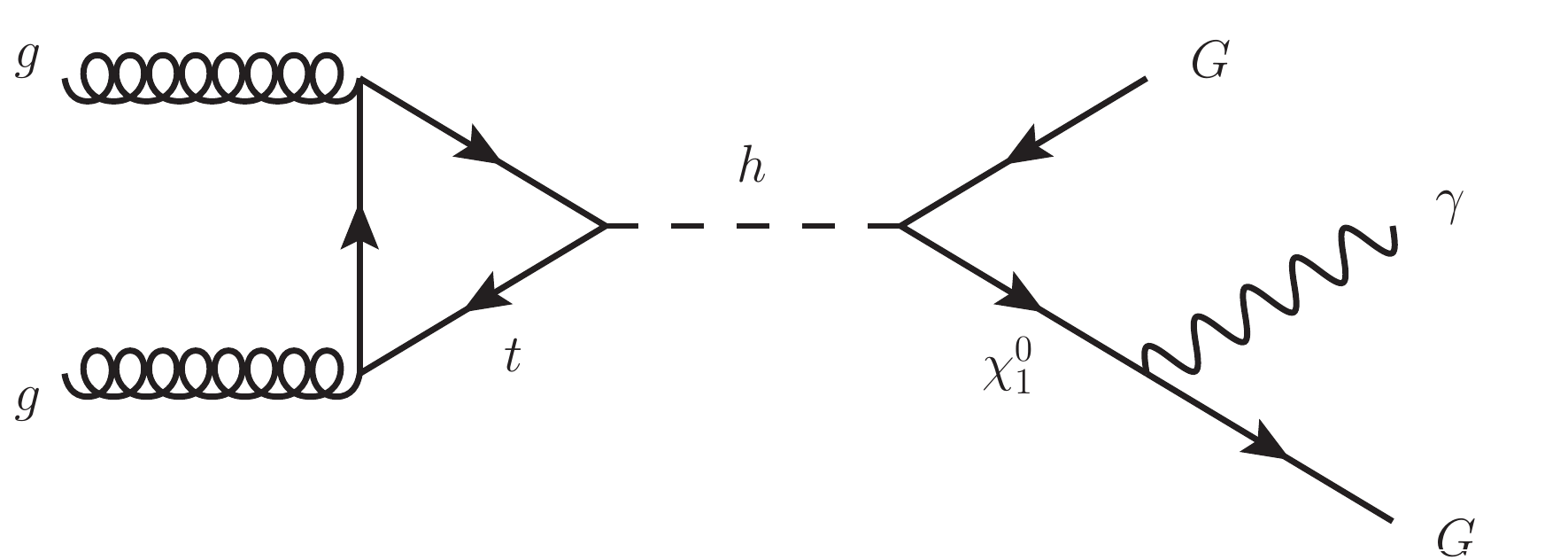}
\includegraphics[scale=0.45]{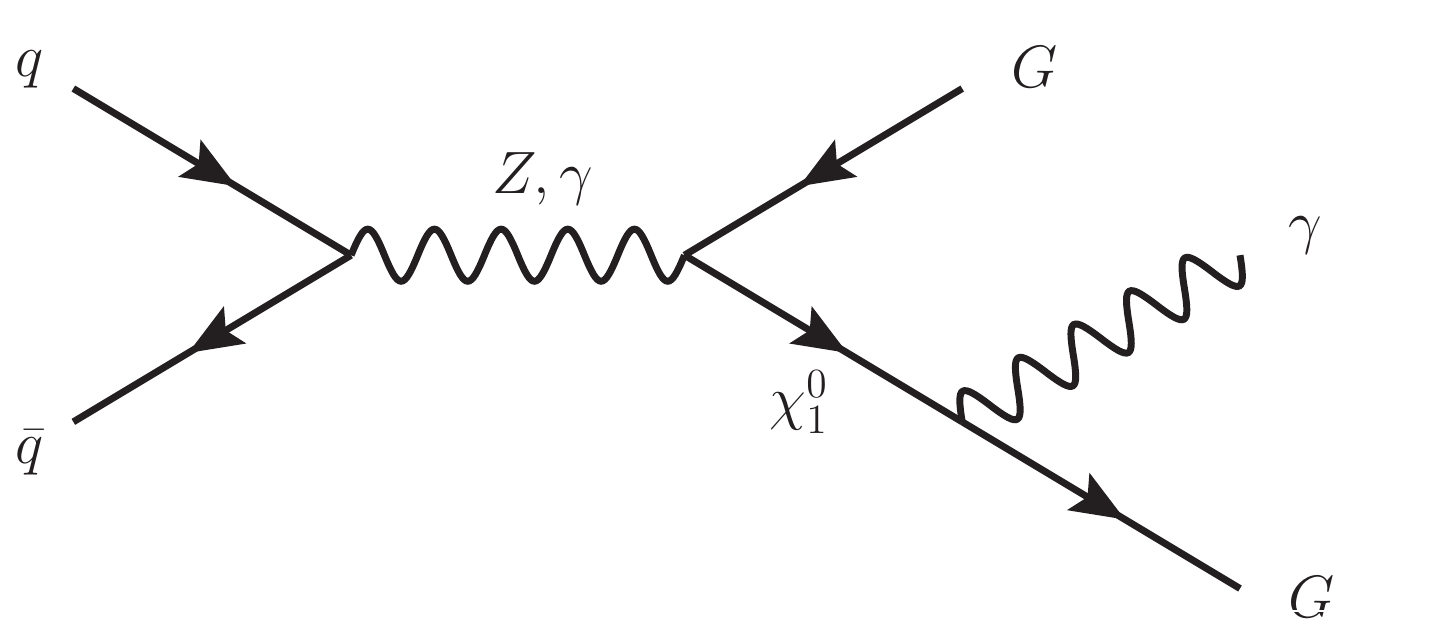}
\end{center}
\caption{ 
\small
Feynman diagrams corresponding to the resonant NP contributions to the final state with a photon$\,+\,\MET\,$in the simplified model in Eq.~\eqref{ph2}.
}\label{Fig1}
\end{figure}

\begin{figure}[!t]
\begin{center}
\includegraphics[scale=0.45]{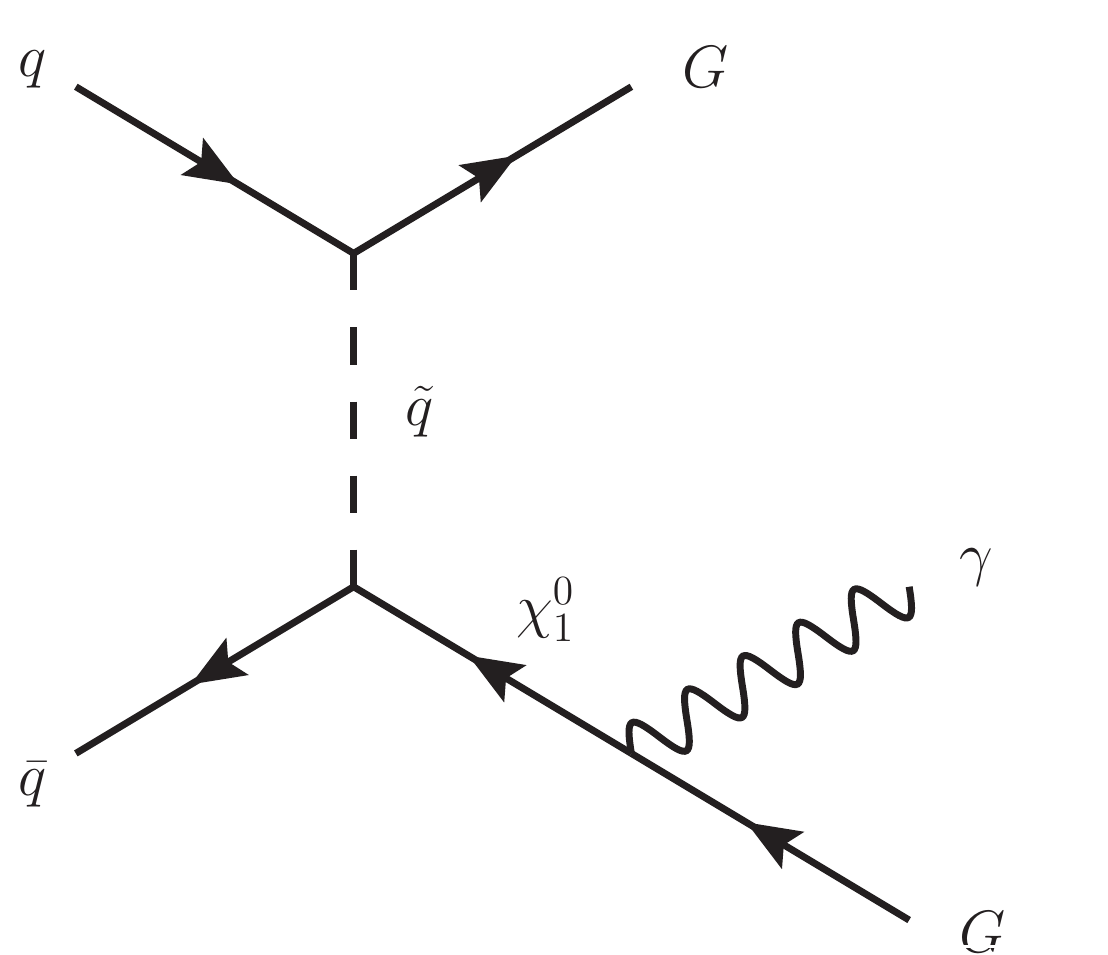}
\end{center}
\caption{ 
\small
Feynman diagram corresponding to the $t$-channel squark exchange contribution to the final state with a photon$\,+\,\MET\,$.
}\label{Fig1b}
\end{figure}

The relevant signal processes contributing to the final state with one isolated photon$\,+\,\MET\,$arising from the Lagrangian \eqref{ph2} are diagrammatically depicted in Fig.~\ref{Fig1}.
The contribution arising from the $s$-channel photon exchange is irrelevant with respect to the others and therefore, it will not be considered in the remainder of this section. 
The process involving an on-shell $Z$ boson is relevant only for $m_{\chi}<m_{Z}$, where it is kinematically allowed. Notice that the $Z$ boson diagram in Fig.~\ref{Fig1} gives rise to photons with lower transverse momenta ($p_{T}^{\g}<m_{Z}/2$) than the one involving the Higgs boson, for which $p_{T}^{\g}<m_{h}/2$. 
Also note that the decay of the neutralino is always prompt for our choice of the relevant parameters. This can be seen from the decay length of the neutralino (produced with energy $E$),  given by
\be\l{ph2b}
L_{\chi}= \f{1}{g_{\chi\gamma}^{2}}\f{\(100\,\text{GeV}\)^{5}}{m_{\chi}^{3}m^{2}}\(\f{\sqrt{F}}{1\,\text{TeV}}\)^{4}\sqrt{\(\f{E^{2}}{m_{\chi}^{2}}-1\)}\,\cdot 10^{-10}\,\text{cm}\,.
\ee
One process which does not arise from the simplified model in Eq.~\eqref{ph2} but which in general contributes to the photon$\,+\,\MET\,\,$channel in low scale SUSY breaking models is the non-resonant process involving a squark $t$-channel exchange, diagrammatically depicted in Fig.~\ref{Fig1b}.  This process, studied for example in Ref.~\cite{Brignole:1998eg},  
is dominant with respect to the resonant processes in Fig.~\ref{Fig1} in the region where $p_{T}^{\g}>m_{h}/2$, i.e. above the kinematic end-points of the $p_{T}^{\g}$ distributions of the processes in Fig.~\ref{Fig1}. In contrast, in the region $p_{T}^{\g}<m_{h}/2$, such a non-resonant process generically turns out to be sub-leading with respect to the processes in Fig.~\ref{Fig1}.\footnote{We have verified that the process in Fig.~\ref{Fig1b} is negligible with respect to the  processes in Fig.~\ref{Fig1} for  $p_{T}^{\g}<m_{h}/2$ in the region of the parameter space studied in Section \ref{Sec:Model}.}  In what follows, we will only consider the region $p_{T}^{\g}<m_{h}/2$ and the processes in Fig.~\ref{Fig1}.

\subsection{Signal and background estimates}

The expected number of events with one isolated photon$\,+\,\MET\,$in the region of interest does not favor the search for a bump in some kinematic distribution. In contrast, it suggests a counting experiment in which the total number of signal and background events are compared in a particular window of the phase space, chosen in order to optimize the signal significance. Since we are only considering the Higgs and $Z$ boson $s$-channel diagrams in Fig.~\ref{Fig1}, we can use the Narrow Width Approximation (NWA) to estimate the number of signal events,
\be\l{ph3}
\bry{l}
N_{\text{sig}}^{h}=\s_{h}^{\text{SM}}\times \text{BR}(h\to \chi_{1}^{0}G)\times \text{BR}(\chi_{1}^{0}\to \g G)\times \mathcal{A}_{\text{sign}}^{h}\times \epsilon_{\g} \times L\,,\\
N_{\text{sig}}^{Z}=\s_{Z}^{\text{SM}}\times \text{BR}(Z\to \chi_{1}^{0}G)\times \text{BR}(\chi_{1}^{0}\to \g G)\times \mathcal{A}_{\text{sign}}^{Z}\times \epsilon_{\g} \times L\,,\\
\ery
\ee 
where $\s_{h}^{\text{SM}}$ and $\s_{Z}^{\text{SM}}$ are the SM Higgs and $Z$ bosons production cross sections, $\text{BR}(h\to \chi_{1}^{0}G)$, $\text{BR}(Z\to \chi_{1}^{0}G)$ and $\text{BR}(\chi_{1}^{0}\to \g G)$ are the BRs involved in the decays, $\mathcal{A}_{\text{sign}}^{h}$ and $\mathcal{A}_{\text{sign}}^{Z}$ are the acceptances to the kinematic cuts, $\epsilon_{\g}$ is the photon identification efficiency and $L$ is the integrated luminosity. It is important to take into account that the NWA is not reliable close to the kinematic threshold of the decay. In our case we have verified that for $m_{h}=125$ GeV the NWA is reliable for neutralino masses $m_{\chi}\lesssim 110$ GeV. In order to present unified results, for neutralino masses larger than this value we will use an effective BR (that we will nevertheless continue to call BR) defined in the following way,
\be\l{ph3}
\text{BR}(h\to \chi_{1}^{0}G)=\f{\s(pp\to \chi_{1}^{0}G)}{\s(pp\to h)}\,.
\ee
Note that we will only use the NWA to set the limits on the BRs. All calculations performed to optimize the cuts and to compute the signal acceptances are done using the explicit $2\to3$ matrix elements.

The Leading Order (LO) Higgs boson production cross section in the SM at the $7(8)$ TeV LHC for $m_{h}=125$ GeV is $\s_{h}^{\text{LO}}=7.44(9.39)$ pb\footnote{Computed with MadGraph5 \cite{Alwall:2011uj} using the CTEQ6M parton distribution functions.}. It is known that higher order corrections are large and positive and for the same mass at Next-to-Next-to LO (NNLO) with Next-to-Next-to Leading Logarithm (NNLL) resummation we get $\s_{h}^{\text{NNLO}}=15.31(19.49)$ pb\footnote{Computed using \href{http://theory.fi.infn.it/cgi-bin/higgsres.pl}{http://theory.fi.infn.it/cgi-bin/higgsres.pl}
with exact top mass dependence up to NLO+NLL, $m_{t}=173.3$ GeV, $m_{b}=4.67$ GeV, $\mu_{r}=\mu_{f}=m_{h}$ and NNLO MSTW2008 parton distribution functions.}. It is also known that the higher order corrections affect the kinematics of the process, since they introduce additional jets which can give a non-negligible $p_{T}$ to the on-shell Higgs (for an updated discussion of $p_{T}$ distributions of Higgs production at higher orders see, e.g., Ref.~\cite{deFlorian:2011uo}). Due to the complexity of the analysis involved when taking into account the Higgs momentum and the additional jets arising due to the higher order corrections, we will  continue to assume that the Higgs is produced at rest even though  we use the NNLO production cross section\footnote{In fact, we generate our signal at LO and use the $k$-factor $k=2.08$ to get the NNLO.} since we expect it to be a more correct estimate for the total rate\footnote{In Ref.~\cite{deFlorian:2011uo} it is shown that the typical Higgs $p_{T}$ is around $10$ GeV and the largest contribution to the production cross section comes from the region $p_{T}^{h}\lesssim 40$ GeV. This means that the main contribution to the radiative corrections come from soft jets which do not affect the selection of the final state we are interested in.}.
For what concerns the $Z$ production cross section we have $\s_{Z}^{\text{LO}}=23.2(27.1)$ nb for the $7(8)$ TeV LHC\footnote{Computed with MadGraph5 using the CTEQ6M parton distribution functions.}. Also the $Z$ production cross section is sensitive to radiative corrections (although less than for $h$) and at NNLO we get $\s_{Z}^{\text{NNLO}}=28.5(33.3)$ nb\footnote{The value of the cross section at $7$ TeV corresponds to the NNLO $Z$ production cross section computed with the MSTW08 parton distribution functions in Ref.~\cite{Watt:2011ev}. The value at $8$ TeV has been obtained by rescaling the value at $7$ TeV by the factor $\s_{Z}^{\text{LO}}(8\text{ TeV})/\s_{Z}^{\text{LO}}(7\text{ TeV})$.}.

The search for a photon$\,+\,\MET\,$signal at low $p_{T}^{\g}$ is difficult due to the fact that the SM background is large in this region and subject to large systematic uncertainties. The main backgrounds to this signal are given by the processes listed in Table \ref{table1}. 
\begin{table}[t]
\begin{center}
\begin{tabular}{c|c|c}
{\bf Name} & {\bf Process} 				& {\bf Detection}			\\\hline
bg1		& $pp\to \g Z \to \g 2\nu$			& $\gamma Z$ irreducible background		\\
bg2		& $pp\to jZ \to j2\nu$				& Jet fakes a photon			\\
bg3		& $pp\to W \to e\nu$				& Electron fakes a photon		\\
bg4		& $pp\to \g j$				& Missing jet				\\
bg5		& $pp\to \g W \to \g l\nu$	& Missing lepton			\\
bg6		& $pp\to \g\g$			& Missing photon			%
\end{tabular}
\end{center}
\caption{\small Main SM contributions to the $\g+\MET$ final state. 
}\l{table1}
\end{table}
The relative importance of these backgrounds depends on the region of $p_{T}^{\g}$ under consideration. In particular, bg1 and bg2 are the most important backgrounds in the high $p_{T}^{\g}$ region, i.e.~$p_{T}^{\g}\gtrsim 90$ GeV (see, e.g., Ref.~\cite{CMS-PAS-EXO-11-058}), while bg3 is only relevant for $p_{T}^{\g}\lesssim m_{W}/2$. The background bg4 is negligible with respect to the others in the region $p_{T}^{\g}\gtrsim 90$ GeV but it turns out to be the dominant background for \mbox{$p_{T}^{\g}\lesssim 70$ GeV}. This last background is subject to many experimental effects like the jet reconstruction efficiency and energy resolution in the forward region. Since we are not going to perform a detector level analysis and since our simulations are done at the parton level, we will simply assume that jets in the very forward region, with a pseudorapidity $|\eta|>4$, are not reconstructed and, when produced in association with an isolated photon, constitute a background for the $\g+\MET$ final state.\footnote{This assumption has been used to compute the $\g+j$ background in the kinematic regime considered in the analysis of Ref.~\cite{CMS-PAS-EXO-11-058}, giving the same result. Note that jet reconstruction in the forward region is a difficult experimental task, especially in presence of high pile-up and only jets with a $p_{T}$ of at least $30\div40$ GeV can be safely reconstructed. For issues related to the jet reconstruction in the very forward region, we refer the reader to the ATLAS and CMS TDR's of Refs.~\cite{ATLASCollaboration:2008ut} and \cite{CMSCollaboration:2006p116} respectively.} The backgrounds bg5 and bg6 are sub-leading in the region of $p_{T}^{\g}$ under consideration and will not be considered in the remainder of this paper.

The number of events corresponding to the backgrounds bg1-bg4 can be written in the following  form,
\be\l{ph5}
\bry{l}
N_{\text{bg1}}=\s_{\text{bg1}}\times \mathcal{A}_{\text{bg1}}\times \epsilon_{\g} \times L\,,\\
N_{\text{bg2}}=\s_{\text{bg2}}\times \mathcal{A}_{\text{bg2}}\times \epsilon_{\g} \times r_{\g-j}^{-1}\times L\,, \\
N_{\text{bg3}}=\s_{\text{bg3}}\times \mathcal{A}_{\text{bg3}}\times \epsilon_{\g} \times r_{\g-e}^{-1}\times L\,, \\
N_{\text{bg4}}=\s_{\text{bg4}}\times \mathcal{A}_{\text{bg4}}\times \epsilon_{\g} \times L\,,
\ery
\ee 
where $r_{\g-j}$ and $r_{\g-e}$ are the rejection factors for the misidentification of a jet and an electron with a photon, respectively. We take the former to be $r_{\g-j}=10^{3}$ for photons with $p_{T}^{\g}>25$ GeV, which is conservative with a photon reconstruction efficiency $\epsilon_{\g}$ larger than $95\%$ (see, e.g.,
Ref.~\cite{ATLASCollaboration:2008ut}), and the latter to be $r_{\g-e}=200$, estimated\footnote{Our estimate is an average in the whole range $p_{T}^{\g}>95$ GeV, but we expect better performances for lower $p_{T}^{\g}$ due to the higher efficiency of electron/photon identification.} from the CMS analysis in Ref.~\cite{CMS-PAS-EXO-11-058}.  

\subsection{Kinematic cuts}

In order to optimize the signal over background ratio $\mathcal{R}=N_{S}/N_{B}$ and the signal significance $\mathcal{S}=N_{S}/\sqrt{N_{B}}$ we need to choose a proper window of the phase space by applying kinematic cuts. Since the $p_{T}^{\g}$ distributions for the signal arising from the processes in Fig.~\ref{Fig1} have an end-point at $m_{h}/2$ and $m_{Z}/2$, respectively, we  choose the upper cut to be $p_{T}^{\g}|_{\text{max}}=m_{h}/2$. The choice of the lower cut is more delicate since backgrounds in this region are both large and difficult to estimate. In order to have an understanding of the sensitivity of the background, the signal and the signal significance to the lower cut $p_{T}^{\g}|_{\text{min}}$, we have computed them for different values of $p_{T}^{\g}|_{\text{min}}$ varying from $30$ GeV to $50$ GeV.
\begin{table}[!t]
\begin{center}
\begin{tabular}{c|c|c|c|c|c}
$p_{T}^{\g}|_{\text{min}}$ & bg1	& bg2 & bg3 & bg4 
& Total Background   \\\hline
$30$		& $742$	& $257$	& $12.2\cdot 10^{3}$	& $14.2\cdot 10^{3}$	& $27.4\cdot 10^{3}$	\\
$35$		& $508$	& $185$	& $8342$		& $6477$		& $15.5\cdot 10^{3}$	\\
$40$		& $341$	& $131$	& $1932$		& $3135$		& $5539$	\\
$45$		& $223$	& $88$	& $157$		& $1507$		& $1975$	\\
$50$		& $137$	& $56$	& $59$		& $690$		& $942$
\end{tabular}
\end{center}
\caption{\small\label{kincuts} The number of background events for the backgrounds bg1-bg4 and the total number of background events (sum of bg1-bg4) for 1 fb$^{-1}$ of integrated luminosity for the kinematic cuts $p_{T}^{\g}|_{\text{min}}<(p_{T}^{\g},p_{T}^{j},p_{T}^{e})<m_{h}/2$, $|\eta_{\g}|,|\eta_{j}^{\text{bg2}}|,|\eta_{e}^{\text{bg3}}|<1.44$, $|\eta_{j}^{\text{bg4}}|>4$ and for $\epsilon_{\g}=0.85$. All the backgrounds have been computed using MadGraph5 \cite{Alwall:2011uj}. The background bg3,  arising from on-shell $W$ production, has been corrected using the NNLO $W$ production $k$-factor obtained by the ratio of the MadGraph5 cross section and the NNLO one computed in Ref.~\cite{Watt:2011ev}, rescaled to $\sqrt{s}=8$ TeV.
}\l{table2}
\end{table}
\begin{table}[!t]
\begin{center}
\begin{tabular}{c|c|c|c||c|c}
$p_{T}^{\g}|_{\text{min}}$ & $N_{\text{sig}}^{h}$ & $N_{\text{sig}}^{Z}$ & $N_S$  & $N_{\text{sig}}^{h}/\sqrt{N_{B}}$ & $N_{S}/\sqrt{N_{B}}$ \\\hline
$30$		& $138$	& $36$		& $174$	 &$3.7$		& $4.7$	\\
$35$		& $107$	& $25$		& $132$    &$3.8$		& $4.7$	\\
$40$		& $80$	& $11$		& $91$	 &$4.8$		& $5.5$	\\
$45$		& $55$	& $2$		& $57$	 &$5.5$		& $5.7$	\\
$50$		& $33$	& $1$		& $34$	 &$4.8$		& $5.0$
\end{tabular}
\end{center}
\caption{\small The number of signal events $N_{\text{sig}}^{h}$ and $N_{\text{sig}}^{Z}$ and the total number of signal events $N_S$ for 1 fb$^{-1}$ of integrated luminosity and expected signal significance at LHC at $8$ TeV with $20$ fb$^{-1}$ for the kinematic cuts $p_{T}^{\g}|_{\text{min}}<p_{T}^{\g}<m_{h}/2$, $|\eta_{\g}|<1.44$ and for $\epsilon_{\g}=0.85$. The total backgrounds $N_{B}$ are given in Table \ref{table2}.
}\l{table3}
\end{table}

In Tables \ref{table2} and \ref{table3} we show the number of background and signal events, respectively, for 1 fb$^{-1}$ of integrated luminosity with different lower $p_{T}^{\g}$ cuts. In Table  \ref{table3} we also give the corresponding expected significance for $20$ fb$^{-1}$ of integrated luminosity. We have assumed a photon reconstruction efficiency $\epsilon_{\g}=0.85$ which is an average value for prompt isolated photons in this region of $p_{T}^{\g}$ \cite{ATLASCollaboration:2010hx}.
The signals have been computed using the CalcHEP matrix element generator \cite{Pukhov:1999p1261,Pukhov:2004ca,Belyaev:Df-Kj9yc} and the corresponding choice of parameters in the simplified model \eqref{ph2} are $m_{\chi}=80$ GeV, $\text{BR}(h\to \chi_{1}^{0}G)=2\cdot 10^{-2}$ and $\text{BR}(Z\to \chi_{1}^{0}G)= 5\cdot 10^{-6}$, where we have used the SM Higgs and $Z$ boson total widths, $\G^{h}_{\text{tot}}=4\times 10^{-3}$ GeV and $\G^{Z}_{\text{tot}}=2.5$ GeV, respectively. Since the dependence of $\mathcal{R}$ and $\mathcal{S}$ on $p_{T}^{\g}|_{\text{min}}$ is insensitive to $\text{BR}(h\to \chi_{1}^{0}G)$ and $\text{BR}(Z\to \chi_{1}^{0}G)$ and only weakly dependent on $m_\chi$, we expect that the value of $p_{T}^{\g}|_{\text{min}}$ which optimizes $\mathcal{R}$ and $\mathcal{S}$ will be robust under variations of the parameters.

In Table \ref{table2} we see the importance of the four leading backgrounds in the relevant region of $p_{T}^{\g}$. The backgrounds bg1 and bg2, which are the most important ones at high $p_{T}^{\g}$, are sub-leading in this region with respect to bg4. The background bg3 can be kept under control with a minimum $p_{T}^{\g}$ cut around or above the \mbox{$m_{W}/2\approx40$ GeV} threshold. For $p_{T}^{\g}|_{\text{min}}\geqslant 40$ GeV, bg4 is the dominant background in the region of interest. 
From Table \ref{table3} we see that the highest signal significance $\mathcal{S}$ corresponds to $p_{T}^{\g}|_{\text{min}}=45$ GeV and therefore we choose this value for our analysis. We can summarize our kinematic requirements as follows,\footnote{The value of the $\eta$ cut correspond to the fiducial region of the CMS barrel ECAL. Notice that the $\eta$ coverage of the ATLAS and CMS barrel ECALs is very similar, respectively $|\eta|<1.475$ \cite{ATLASCollaboration:2008ut} and $|\eta|<1.479$ \cite{CMSCollaboration:2006p116}, motivating our choice for this cut.}
\be\l{ph6}
45\text{ GeV }<p_{T}^{\g}<m_{h}/2~, \qquad\qquad |\eta_{\g}|<1.44\,.
\ee
As can be seen in Table \ref{table3}, for this choice of the cuts where $p_{T}^{\g}|_{\text{min}}\approx m_{Z}/2$, the contribution to the signal arising from the resonant $Z$ process in Fig.~\ref{Fig1} is negligible with respect to the resonant Higgs process. Therefore, in what follows, we will focus only on the Higgs process.

The only remaining ingredient that we need for studying the sensitivity of the LHC to $\text{BR}(h\to \chi_{1}^{0}G)$ as a function of the neutralino mass is the signal acceptance to the cuts of Eq.~\eqref{ph6} for the different neutralino masses. These acceptances, shown  in Table \ref{table4} for neutralino masses in the range $60<m_{\chi}<120$ GeV, depend on the Lorentz structure of the relevant interactions but not  on the production rate or the BR.
\begin{table}[t]
\begin{center}
\begin{tabular}{c|c}
$m_{\chi_{1}^{0}}$ (GeV)  	& $\mathcal{A}_{\text{sign}}^{h}$ 	\\\hline

$60$		& $0.126$	\\
$70$		& $0.141$	\\
$80$		& $0.165$	\\
$90$		& $0.198$	\\
$100$	& $0.262$	\\
$110$	& $0.370$	\\
$120$	& $0.418$	
\end{tabular}
\end{center}
\caption{\small\label{kincuts} Signal acceptances to the kinematic cuts in Eq.~\eqref{ph6} for the signal process $pp\to h\to \chi_{1}^{0}G\to  \gamma GG$. 
}\l{table4}
\end{table}

\subsection{Discovery and exclusion limits}

\begin{figure}[!t]
\begin{center}
\includegraphics[scale=0.36]{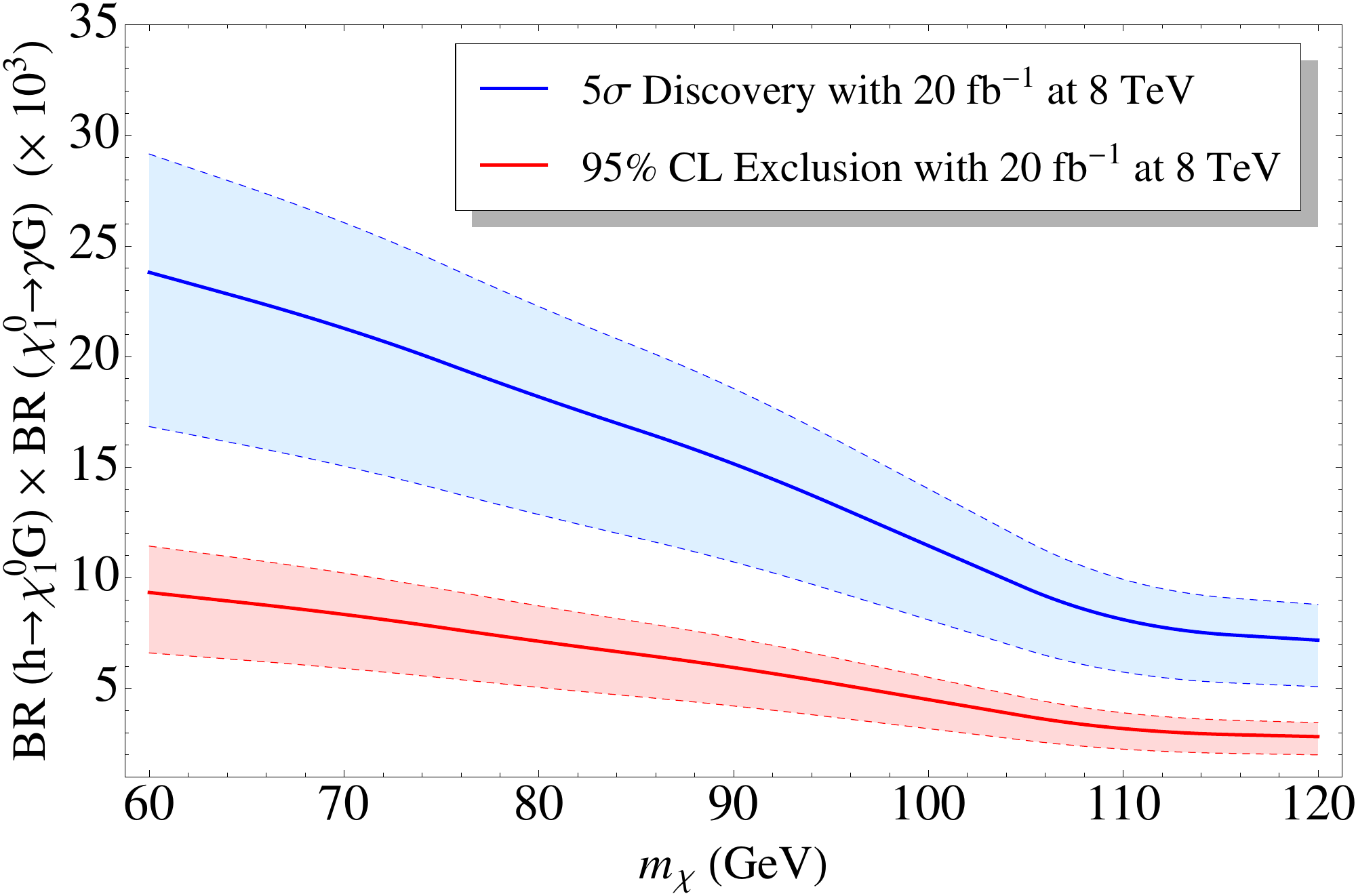}\hspace{2mm}
\includegraphics[scale=0.36]{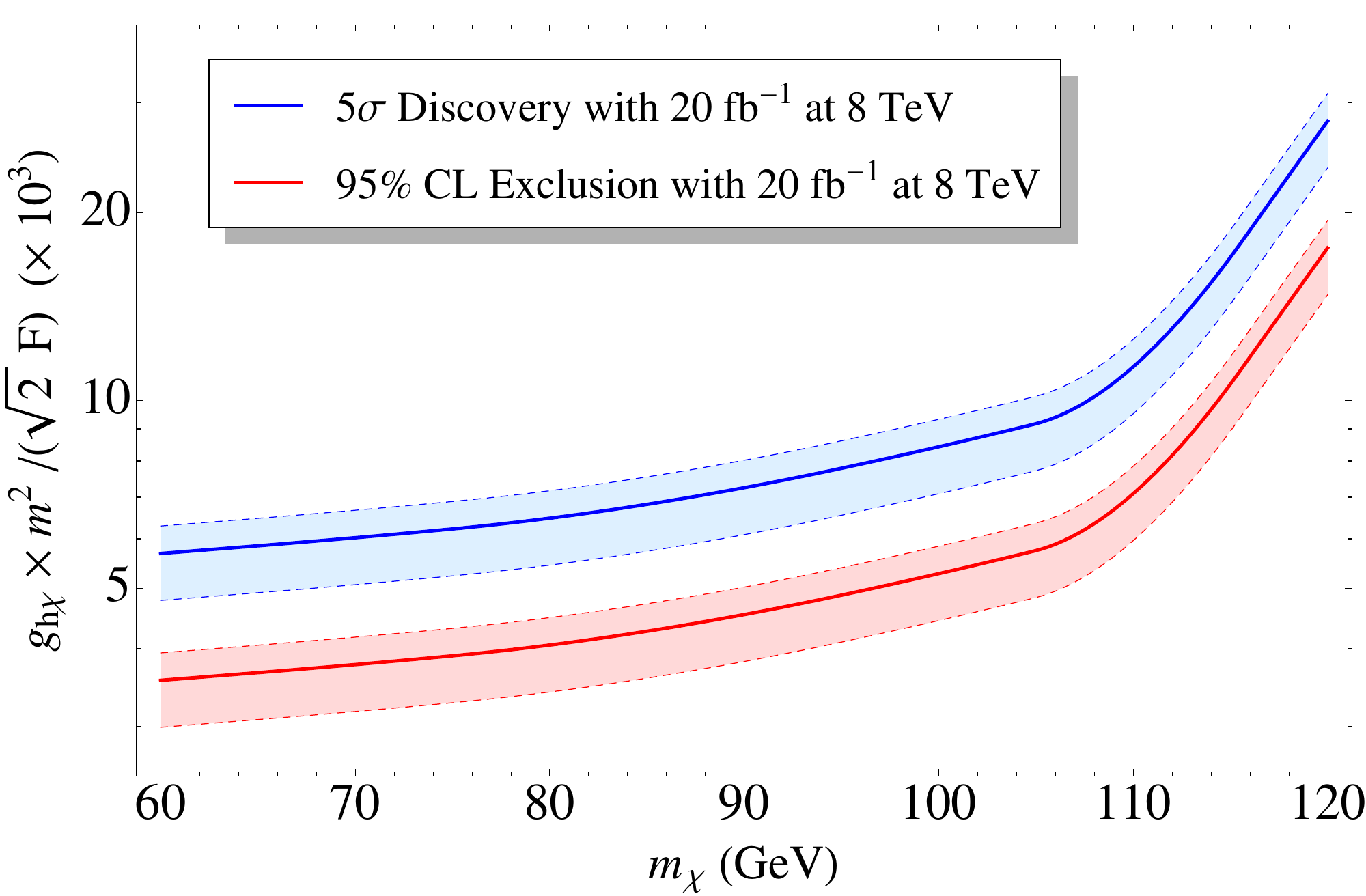}
\end{center}
\caption{ 
\small
Sensitivity of the LHC at $8$ TeV to the quantity $\text{BR}(h\to \chi_{1}^{0}G)\times \text{BR}(\chi_{1}^{0}\to \g G)$ (left panel) and Higgs-neutralino-goldstino coupling (right panel) with $20$ fb$^{-1}$. The region above the solid blue line can be discovered with a $5\s$ significance while the region above the red solid line can be excluded at $95\%$ CL. The bands show the sensitivity of the limits to a $50\%$ variation of the background. The behavior of the curves for $m_{\chi}\gtrsim 110$ GeV is due to the deviation from the NWA close to the threshold of the decay $h\to \chi_{1}^{0}G$.
}\label{Fig2}
\end{figure}

We are now ready to compute the minimum value of $\text{BR}(h\to \chi_{1}^{0}G)$ that can be discovered/excluded at the LHC at $8$ TeV with a given integrated luminosity. In order to be as general as possible, we set a limit on \mbox{$\text{BR}(h\to \chi_{1}^{0}G)\times \text{BR}(\chi_{1}^{0}\to \g G)$} since the latter BR can be smaller than one for $m_{\chi}>m_{Z}$. The limit on this product with a given significance $\mathcal{S}$ and integrated luminosity $L$ is given by,\footnote{This formula has been obtained assuming Gaussian statistics for the background.}
\be\l{ph7}
\Big[\text{BR}(h\to \chi_{1}^{0}G)\times \text{BR}(\chi_{1}^{0}\to \g G)\Big]_{\text{min}}=\f{\mathcal{S}\sqrt{N_{B}}}{\sigma_{h}^{\text{NNLO}}\times \mathcal{A}_{\text{sign}}^{h}\times\epsilon_{\g}\times L}\,.
\ee
Using this relation,  the $95\%$ CL exclusion and the $5\s$ discovery limits are shown in the left panel of Fig.~\ref{Fig2}. In the right panel of the same figure we also show the sensitivity to the Higgs-neutralino-goldstino coupling in the effective Lagrangian \eqref{ph1}.
The limit on the coupling has been obtained by using the expression of the partial width in Eq.~\eqref{ph3a} as a function of the coupling. From this expression we can obtain a limit on the coupling as a function of the product of BRs appearing in the left panel of Fig.~\ref{Fig2},
\be\l{ph7a}
\dst \left[\f{g_{h\chi}m^{2}}{\sqrt{2}F}\right]_{\text{min}}=\sqrt{\f{\Big[\text{BR}(h\to \chi_{1}^{0}G)\times \text{BR}(\chi_{1}^{0}\to \g G)\Big]_{\text{min}}}{\text{BR}(\chi_{1}^{0}\to \g G)}\(1-\f{m_{\chi}^{2}}{m_{h}^{2}}\)^{-2}\f{8\pi\G^{h}_{\text{tot}}}{m_{h}}}\,.
\ee
In the right panel of Fig.~\ref{Fig2} we have assumed  $\text{BR}(\chi_{1}^{0}\to \g G)=1$ but it is simple to rescale the limit for different values of $\text{BR}(\chi_{1}^{0}\to \g G)$ by using Eq.~\eqref{ph7a}.
Notice from Eq.~\eqref{ph7}, by taking into account that the number of background events scales with the photon efficiency $\epsilon_{\g}$, it follows that the limit on the BRs in Eq.~\eqref{ph7} scales as $\epsilon_{\g}^{-1/2}$ while the limit on the coupling in Eq.~\eqref{ph7a} scales as $\epsilon_{\g}^{-1/4}$. Here we have assumed $\epsilon_{\g}=0.85$ but Fig.~\ref{Fig2} can be easily generalized for different values of $\epsilon_{\g}$.

As a final comment we should stress the fact that our analysis has been performed at patron level and without taking into account any systematic uncertainty. The signal over background ratio $\mathcal{R}$ for the signal giving rise to a significance $\mathcal{S}$, considering only statistical error, is given by,
\be\l{ph8}
\mathcal{R}
=\f{\mathcal{S}}{\sqrt{N_{B}}}\,,
\ee
which means $N_{S}/N_{B}=4.4\%(11.3\%)$ for $\mathcal{S}=1.96(5)$ at the LHC at $8$ TeV with $20$ fb$^{-1}$. This implies that a strong knowledge of the systematic uncertainties is necessary and that dedicated experimental studies are needed in order to estimate the backgrounds at the required level of precision. From our estimate of the sensitivity of the limits to a $50\%$ variation of the background, shown by the bands in Fig.~\ref{Fig2}, we conclude that couplings above $10^{-3}\div10^{-2}$ can lead to a discovery  by the 2012 LHC run.

\section{Effective goldstino interactions}\l{Sec:Model}

In this section we discuss how the parameters of the simplified model in the previous section can be related to the parameters of the MSSM. Since all the four vertices in Eq.~\eqref{ph2} involve linear couplings to the neutralino and the goldstino, we will be interested in the lowest order interactions between the goldstino and the MSSM fields. These  interactions can be obtained by coupling the goldstino either derivatively to the supercurrent or, upon integration by parts and using the equations of motion, non-derivatively to the divergence of the supercurrent. Such a divergence is proportional to the mass splitting inside the MSSM supermultiplets, implying that the strength of the interactions are determined by ratios of the MSSM soft parameters over the SUSY breaking scale $f$. 

In Figure \ref{fig:widthBRs} we have set the mass scale $m^{2}=m_{h}^{2}-m_{\chi}^{2}$ to give an idea of what the BR of the Higgs into a neutralino and a goldstino could be in the simplest case where the lightest neutralino is purely higgsino. In this section we are interested in going beyond this simple case by considering mixing in the neutralino mass matrix. The main goal of this section is to show that having a lightest neutralino not completely Higgsino can increase the $\text{BR}\(h\to \chi_{0}^{1}G\)$ leading to an early discovery of this new Higgs decay at the LHC.

The relevant goldstino interactions can be derived from an effective Lagrangian with manifest, but spontaneously broken, SUSY in which the MSSM soft terms have been promoted to supersymmetric operators  involving a goldstino superfield and the MSSM superfields. Since the prefactor of these operators will be given by ratios of soft parameters over $f$, in order to have a reliable effective description we require $\sqrt{f}$ to be larger than any soft parameter. On the other hand, a large contribution to the Higgs-goldstino-neutralino coupling implies a small separation between $\sqrt{f}$ and the soft parameters. As we will see below, in order for the Higgs decay channel under consideration to be relevant for the 2012 LHC run, if the soft parameters are around the TeV scale, $\sqrt{f}$ is required to be in the few TeV region. Effective Lagrangians with manifest supersymmetry with such a low $\sqrt{f}$ have been discussed, for example, in Refs.~\cite{1997hep.ph...11516B,Brignole:1998eg,Brignole:2003hb,Antoniadis:2007et}. Moreover, in Ref.~\cite{Gherghetta:2011tn} it was discussed how to achieve a viable superpartner spectrum in this scenario.

In this section we will not consider any specific mechanisms for SUSY breaking or mediation, but instead discuss how contributions to the vertices in Eq.~\eqref{ph2} can arise from the set of SUSY operators considered in \mbox{Refs.~\cite{Komargodski:2009cq,Antoniadis:2010hs}}. In contrast to the more conventional way of parametrizing spontaneous SUSY breaking, where a background spurion field is introduced containing only a constant auxiliary component, it was prescribed in Ref.~\cite{Komargodski:2009cq} to replace the spurion by a non-linear superfield, $X_{\mathrm{nl}}=\psi_X \psi_X/(2F_X)+\sqrt{2}\theta \psi_X +\theta^2 F_X$, where $\psi_X$ becomes the goldstino at low energies and $F_X$ is the dynamical auxiliary field that acquires a non-vanishing VEV and breaks SUSY. This prescription was applied in Ref.~\cite{Antoniadis:2010hs} to the MSSM, where each of the MSSM soft terms was promoted to a supersymmetric operator. These operators can be obtained by simply multiplying each term in the supersymmetric part of the MSSM by $X$ if the term is holomorphic and $X^\dagger X$ if the term is non-holomorphic. 
Any operator containing additional powers of the goldstino superfield would automatically vanish since the  non-linear goldstino superfield satisfies $X_{\mathrm{nl}}^2=0$ \cite{Komargodski:2009cq}.


\subsection{Couplings relevant for the Higgs decay}

In order to relate the prefactor of the first term in Eq.~\eqref{ph2} to MSSM parameters we start by considering supersymmetric operators which give rise to Higgs-goldstino-neutralino couplings. One example of such an operator is obtained by promoting the down and up-type Higgs scalar soft terms to the following SUSY operators,           
\begin{eqnarray}
\label{LkinX}
\sum_{I=d,u}-\int d^4 \theta \, \frac{m_{I}^2}{f^2}\, X_{\mathrm{nl}}^\dagger X_{\mathrm{nl}}  \, H_{I}^{\dagger} e^{\,gV} H_{I} \supset -\sum_{I=d,u} \frac{m_{I}^2}{f}\psi_X \psi_{H_I^0} h_I^0{}^\ast  +\mathrm{h.c.} ~,
\end{eqnarray}
where we have used that the auxiliary field of the goldstino supermultiplet acquires a non-vanishing VEV, $F_X^\dagger=-f+\ldots$.  In Eq.~\eqref{LkinX} the interactions of the MSSM vector supermultiplets with the Higgs doublets are represented by the factor $e^{\,gV}$, with $V$ in the appropriate representation of the SM gauge group and with the corresponding gauge coupling constants. The MSSM down and up-type Higgs scalar soft terms arise from the operators in Eq.~\eqref{LkinX}, upon extracting the terms bilinear in the goldstino supermultiplet auxiliary fields and inserting their VEVs. Another contribution to the Higgs-goldstino-neutralino coupling is obtained by promoting the $B_\mu$ soft term to the following supersymmetric operator,
\begin{equation}
\label{LBmuX}
-\int d^2 \theta \frac{B_\mu}{f} X_{\mathrm{nl}}  H_d \,H_u +\mathrm{h.c.} \supset \frac{B_\mu}{f} \psi_X \left( \psi_{H_d^0} h_u^0 +  \psi_{H_u^0} h_d^0 \right) +\mathrm{h.c.} ~,
\end{equation}
where $H_d\, H_u=H_{d}^{0} H_{u}^{0}-H_{d}^{-} H_{u}^{+}$.

After EW symmetry breaking (EWSB), there will also be contributions to the Higgs-goldstino-neutralino coupling, for example, arising from the following supersymmetric version of the gaugino soft mass terms, 
\begin{eqnarray}
\label{LgX}
\hspace{-4mm} \sum_{i=1}^{2}\int d^2\theta\,  \frac{m_{i}}{2f}X_{\mathrm{nl}}   \,
W_{A_i}^\alpha W_{\alpha}^{A_i}+ \mathrm{h.c.} \supset -\frac{m_{1}}{\sqrt{2}f} \psi_X D_{1} \lambda_{1} - \frac{m_{2}}{\sqrt{2}f} \psi_X D_{2}^{(3)} \lambda_{2}^{(3)}+ \mathrm{h.c.}\,,
\end{eqnarray}
where the indices $A_1=1$ and $A_2=1,2,3$ run over the adjoint representation of the $U(1)_Y$ and $SU(2)_L$ gauge groups and the $D$-terms for the $U(1)_Y$ and the (third component of) $SU(2)_L$  are given by,
\be\label{Dterms}
\bry{lll}
D_{1}&=&\dst\frac{g_1}{2} \left( |h_d^0|^2- |h_u^0|^2 \right)~, \vspace{2mm}\\
D_{2}^{(3)}&=&\dst-\frac{g_2}{2} \left( |h_d^0|^2- |h_u^0|^2 \right)~.
\ery
\ee
Moreover, after EWSB, the true goldstino does not in general only have a $\psi_X$ component, but also  components from the neutral higgsinos, $\psi_{H_u^0}$ and $\psi_{H_d^0}$, and the neutral gauginos, $\lambda_1$ and $\lambda_2^{(3)}$ due to non-vanishing $F$ and $D$-components of the corresponding superfields. Therefore, contributions to the relevant coupling can arise from the ordinary kinetic terms for the Higgs superfields, 
\begin{eqnarray}
\label{Hkin}
\sum_{I=d,u} \int d^4 \theta  \,H_{I}^{\dagger} e^{\,gV} H_{I}  &\supset& -\frac{g_1}{\sqrt{2}} h_u^0{}^\ast \lambda_1 \psi_{H_u^0} + \frac{g_1}{\sqrt{2}} h_d^0{}^\ast \lambda_1 \psi_{H_d^0} \nn \\
&&+ \frac{g_2}{\sqrt{2}} h_u^0{}^\ast \lambda_2^{(3)} \psi_{H_u^0} -\frac{g_2}{\sqrt{2}} h_d^0{}^\ast \lambda_2^{(3)} \psi_{H_d^0}+ \mathrm{h.c.} \,.
\end{eqnarray}
Let us collect the contributions to the Higgs-neutralino-goldstino coupling we obtained from the supersymmetric operators  considered in Eqs.~\eqref{LkinX}, \eqref{LBmuX}, \eqref{LgX} and \eqref{Hkin}, 
\begin{eqnarray}
\label{higgsvert}
\mathcal{L}_{h\chi G}&=& -\frac{m_u^2}{f}\psi_X \psi_{H_u^0} h_u^0{}^\ast -\frac{m_d^2}{f}\psi_X  \psi_{H_d^0} h_d^0{}^\ast+\frac{B_\mu}{f} \psi_X \left( \psi_{H_d^0} h_u^0 +  \psi_{H_u^0} h_d^0 \right) \nn \\
&&-\frac{m_{1}}{\sqrt{2}f} \psi_X D_{1} \lambda_{1} - \frac{m_{2}}{\sqrt{2}f} \psi_X D_{2}^{(3)} \lambda_{2}^{(3)}   \\
&&-\frac{g_1}{\sqrt{2}} h_u^0{}^\ast \lambda_1 \psi_{H_u^0} + \frac{g_1}{\sqrt{2}} h_d^0{}^\ast \lambda_1 \psi_{H_d^0} + \frac{g_2}{\sqrt{2}} h_u^0{}^\ast \lambda_2^{(3)} \psi_{H_u^0} -\frac{g_2}{\sqrt{2}} h_d^0{}^\ast \lambda_2^{(3)} \psi_{H_d^0}+\mathrm{h.c.} \nn ~.
\end{eqnarray}
As will be evident from the discussion below, the terms appearing in the second and third line of Eq.~\eqref{higgsvert} will in general be small compared to the terms appearing in the first line. The reason is that these contributions are always suppressed by ratios of $v^{2}$ or $v\mu$ over $f$~\footnote{We are taking the $\mu$-parameter to be small, around 100 GeV.}. Note that there are other non-renormalizable supersymmetric operators
which have the same, or even lower, dimension as the ones that we consider, e.g. $(H_dH_u)^2$ in the superpotential or quartic Higgs superfield couplings in the Kahler potential. However, these operators will not give a significant contribution to the Higgs-goldstino-neutralino coupling due to the suppression of $v^{2}$ or $v\mu$ over $f$ ratios.

\subsection{Couplings relevant for the neutralino decay}

The second term in Eq.~\eqref{ph2}, relevant for the neutralino decay into a photon and a goldstino, can arise from the supersymmetric version of the gaugino soft mass term, given in Eq.~\eqref{LgX}, but where the following component interactions are extracted, 
 \begin{eqnarray}
\label{photonvert}
\mathcal{L}_{\chi \gamma G}=  -\frac{1}{\sqrt{2}f} \psi_X \sigma^{\mu\nu} F_{\mu\nu}
\left( m_{1}\lambda_1\cos\theta_w+  m_{2}\lambda_2^{(3)}\sin\theta_w \right) + \mathrm{h.c.}~,
\end{eqnarray} 
where the $U(1)_Y$ and the $SU(2)_L$ gauge field strengths $B_{\mu\nu}$ and $W^{(3)}_{\mu\nu}$ have been rewritten in terms of the photon and $Z$ boson fields,
\be\label{gaugetransf}
\bry{lll}
B_{\mu}&=& \cos\theta_w A_{\mu}- \sin\theta_w Z_{\mu}~, \vspace{2mm}\\
W^{(3)}_{\mu}&=& \sin\theta_w A_{\mu}+ \cos\theta_w Z_{\mu}~,
\ery
\ee
and $\sin\theta_w =g_1/\sqrt{g_1^2+g_2^2}$ and $\cos\theta_w =g_2/\sqrt{g_1^2+g_2^2}$. 

The third term in Eq.~\eqref{ph2} is given by the coupling analogous to Eq.~\eqref{photonvert} of the transverse $Z$ boson components to the goldstino. In addition, the last term in Eq.~\eqref{ph2}, is given by the coupling of the longitudinal $Z$ boson component arising from the kinetic terms for the Higgs superfields in Eq.~\eqref{Hkin}. By collecting both kinds of terms, we get the following contribution to the neutralino-$Z$-goldstino coupling,
 \begin{eqnarray}
 \label{Zvert}
\mathcal{L}_{\chi Z G}&=&  -\frac{1}{\sqrt{2}f} \psi_X \sigma^{\mu\nu} Z_{\mu\nu}
\left( -m_{1}\lambda_1\sin\theta_w+  m_{2}\lambda_2^{(3)}\cos\theta_w \right) \nn \\
 &&-\frac{1}{4}\sqrt{g_1^2+g_2^2}\left( \overline{\psi}_{H_d^0}\,\overline{\sigma}^\mu\, \psi_{H_d^0}-\overline{\psi}_{H_u^0}\,\overline{\sigma}^\mu\, \psi_{H_u^0}  \right)Z_\mu + \mathrm{h.c.}~.
\end{eqnarray}


\vspace{2mm}
\subsection{The mass basis of the Higgs and the neutralinos}

As it was mentioned above, there are other operators, beyond those in Eqs.~\eqref{LkinX} and \eqref{LBmuX}, which, from a strict effective field theory point of view, one should consider, e.g. $(H_dH_u)^2$ or $X_{\mathrm{nl}}(H_dH_u)^2$ in the superpotential or quartic Higgs superfield couplings in the Kahler potential. Since we do not specify the underlying dynamics or symmetries of the SUSY breaking sector, such operators, none of which have dimension higher than 6, can give significant contributions, for instance, to the tree level masses of the Higgs bosons and the neutralinos (see Refs.~\cite{Brignole:2003hb,Dine:2007eg, Antoniadis:2007et, Antoniadis:2008cu, Carena:2009ey, Carena:2011ty, Boudjema:2011un, Boudjema:2012wp} for discussions concerning a more general set of effective operators and how they can affect the MSSM Higgs sector). In this section we do not attempt to make a complete analysis of the most general effective model of low scale SUSY breaking but instead we provide a discussion concerning how the particular operators in Eqs.~\eqref{LkinX} and \eqref{LBmuX} affect the tree level mass of the lightest neutral CP-even Higgs particle and moreover, how the soft parameters appearing in the prefactors of these operators can be related to the couplings of the simplified model in Eq.~\eqref{ph2}. 

Integrating out the $F_X$-component of the goldstino superfield in Eqs.~\eqref{LkinX}, \eqref{LBmuX} and \eqref{LgX} gives rise to quartic Higgs couplings in the tree level F-term scalar potential. These contributions adds to the usual MSSM contribution, originating from the quartic D-term potential, and their combined contribution gives rise to the following tree level mass for the lightest neutral CP-even Higgs particle \cite{Petersson:2011in},
\begin{eqnarray}
\label{mh}
m_{h,\mathrm{tree}}^{2} & = & m_{Z}^2 \cos^2 2\beta+v^2   \left(  \frac{2\mu^2}{f}-\frac{B_\mu}{f} \sin2\beta \right)^2 ~,
\end{eqnarray}
where the first term is the standard MSSM D-term contribution and the second term is the additional F-term contribution. In order to obtain a viable mass spectrum for all the Higgs particles it is necessary to set $B_\mu>0$, implying a destructive interference between the two terms in the parenthesis in Eq.~\eqref{mh}. For us, since we are interested in having an NLSP neutralino with a significant higgsino-fraction, it is natural to take $B_\mu \sin2\beta$ significantly larger than $2\mu^2$  in order to raise the tree level Higgs mass. This implies that, in order for the tree level Higgs mass to receive as large contribution as possible from this term it is favorable to have $\tan\beta$ small, in contrast to the MSSM D-term contribution which is maximized at large $\tan\beta$. This contribution is analogous to the one obtained in the context of the NMSSM 
(see Ref.~\cite{Ellwanger:2009en} for a review), arising from an operator analogous to the marginal one in Eq.~\eqref{LBmuX}, with $(B_\mu/f)$ replaced by a free dimensionless parameter $\lambda$, unrelated to any soft parameter\footnote{For the case of large $\lambda\sim1\div 2$ see Refs.~\cite{2007PhRvD..75c5007B,Hall:2011ty}.}.

We can rewrite all the relevant gauge basis vertices, Eqs.~\eqref{higgsvert}, \eqref{photonvert} and \eqref{Zvert}, in terms of the mass basis ones by using the EWSB vacuum of Ref.~\cite{Petersson:2011in}, in which the supersymmetric operators in Eqs.~\eqref{LkinX}, \eqref{LBmuX}, \eqref{LgX} and \eqref{Hkin} were taken into account.
Since we are only interested in the lightest SM-like mass eigenstate $h$ we only need the following two entries in the rotation matrix $R$ which diagonalizes the $3\times3$ mass matrix for the neutral real scalars,  
\be\label{higgstransf}
\bry{lll}
\mathrm{Re}\,h_d^0&\to& R_{(h,d)}h~, \vspace{2mm}\\
\mathrm{Re}\,h_u^0&\to& R_{(h,u)}h~.
\ery
\ee

In the aforementioned EWSB vacuum, in the canonically normalized gauge eigenbasis $(\lambda_1,\lambda_{2}^{(3)},\psi_{H_d^0},\psi_{H_u^0},\psi_X)$, the neutralino mass matrix is given by, 
\small\begin{equation}
\label{neutralino5b5}
 \left(\begin{array}{ccccc}
 m_1 & 0 & -m_Z \sin\theta_w \cos\beta & m_Z \sin\theta_w \sin\beta & \frac{g_1 v^2 m_1}{2\sqrt{2}f}\cos 2\beta \\
 0 & m_2 & m_Z \cos\theta_w \cos\beta & -m_Z \cos\theta_w \sin\beta & -\frac{g_2 v^2 m_2}{2\sqrt{2}f} \cos 2\beta\\
 -m_Z \sin\theta_w \cos\beta & m_Z \cos\theta_w \cos\beta & 0 & \mu & -\frac{\mu^2 v}{f} \cos\beta\\
 m_Z \sin\theta_w \sin\beta & -m_Z \cos\theta_w \sin\beta & \mu & 0 & -\frac{\mu^2 v}{f} \sin\beta \\
\frac{g_1 v^2 m_1}{2\sqrt{2}f}\cos 2\beta & -\frac{g_2 v^2 m_2}{2\sqrt{2}f} \cos 2\beta& -\frac{\mu^2 v}{f} \cos\beta & -\frac{\mu^2 v}{f} \sin\beta & 0
 \end{array}\right).
\end{equation}\normalsize
Note that, throughout the paper, we do not consider terms suppressed by $1/f^2$ unless they are proportional to the ratio $B_\mu/f$. In order to disentangle the goldstino and to have canonically normalized kinetic terms we perform the following transformations,
\be\label{transf}
\bry{lll}
\lambda_1&\to&\dst \lambda_1-\frac{g_1 v^2 \cos2\beta}{2\sqrt{2}f}G~, \vspace{2mm}\\
\lambda_{2}^{(3)}&\to&\dst \lambda_{2}^{(3)}+\frac{g_2 v^2 \cos2\beta}{2\sqrt{2}f}G~, \vspace{2mm}\\
\psi_{H_d^0}&\to&\dst \psi_{H_d^0}+\frac{\mu\, v \,\sin\beta}{f}G~, \vspace{2mm}\\
\psi_{H_u^0}&\to&\dst \psi_{H_u^0}+\frac{\mu\, v \,\cos\beta}{f}G~, \vspace{2mm}\\
\psi_X&\to&\dst G- \frac{v \,\mu\,\cos\beta}{f}\, \psi_{H_u^0}- \frac{v \,\mu\, \sin\beta}{f}\,\psi_{H_d^0}+ \frac{g_1 v^2 \cos 2\beta}{2\sqrt{2}f}\,\lambda_1- \frac{g_2 v^2 \cos 2\beta}{2\sqrt{2}f} \,\lambda_{2}^{(3)}~,
\ery
\ee
which make all the entries in the fifth row and the fifth column of the mass matrix in Eq.~\eqref{neutralino5b5} vanish. We are therefore left with a  neutralino mass matrix which, in the gauge eigenbasis $(\lambda_1,\lambda_{2}^{(3)},\psi_{H_d^0},\psi_{H_u^0})$, is given by the upper left $4\times4$ block of Eq.~\eqref{neutralino5b5}, that can be diagonalized by a unitary matrix $N$.  
Since we will here only be interested in the lightest neutralino $\chi^{0}_{1}$, we only need the following four entries in the $N$ matrix in order to rewrite the gauge basis fermions,
\be\label{neutralinotransf}
\bry{lll}
\lambda_1&\to&N_{(1,B)}\chi^{0}_{1}~, \vspace{2mm}\\
\lambda_{2}^{(3)}&\to&N_{(1,W)}\chi^{0}_{1}~, \vspace{2mm}\\
\psi_{H_d^0}&\to&N_{(1,d)}\chi^{0}_{1}~, \vspace{2mm}\\
\psi_{H_u^0}&\to&N_{(1,u)}\chi^{0}_{1}~.
\ery
\ee

The $h \chi^0_1 G$-vertex in Eq.~\eqref{ph2} can now be obtained from Eq.~\eqref{higgsvert} by using Eqs.~\eqref{higgstransf}, \eqref{transf} and \eqref{neutralinotransf}. In an analogous way, the $ \chi^0_1 \gamma G$ and $ \chi^0_1 Z G$ vertices in Eq.~\eqref{ph2} can be obtained from Eqs.~\eqref{photonvert} and \eqref{Zvert}. The analytic form of these vertices are given in Appendix \ref{appendix}.

\subsection{Relation to the discovery and exclusion limits}\label{Sec:Sub}

By using the couplings for the $h \chi^0_1 G$, $ \chi^0_1 \gamma G$ and $ \chi^0_1 Z G$ vertices derived from the effective model above, whose analytic forms are given in Appendix \ref{appendix}, we can now analyze the parameter space and explore the possibility of obtaining a photon$\,+\,\MET\,$signal compatible with the discovery and exclusion limits obtained in the simplified model of Section \ref{Sec:Pheno}. In order to have a large $h \chi^0_1 G$-coupling it is necessary for the NLSP neutralino to have a significant higgsino component, implying that $\mu$ should be rather small, of the order of $100$ GeV. For smaller values of $\mu$, there is a tension due to the fact that $\mu$ also appears in the formula for the chargino masses, 
\begin{equation} \label{charginos}
m_{\chi^{\pm}_i}^2 = \frac{1}{2} \left[(m_2^2 + \mu^2+2 m_W^2 \pm \sqrt{(m_2^2 + \mu^2+2 m_W^2)^2- 4(\mu ~m_2 + m_W^2 \sin^2{2 \beta})^2} \right]~.\\
\end{equation} 
Since also the wino mass $m_2$ appears in Eq.~\eqref{charginos}, it is bounded from below in a similar way as $\mu$. In contrast, the bino mass $m_1$ only appears in the neutralino mass matrix, implying that  the  lower region of the neutralino mass range can be reached  by taking  $m_1$ to be lower than or of the same order as $\mu$, even though this implies that the NLSP neutralino will have a significant bino-component. Note that a non-vanishing bino-component is needed in order for the neutralino to decay into a photon and a goldstino. However, as can be seen from Eq.~\eqref{ph2b}, for low $\sqrt{f}$ this decay is always prompt, almost independently of this coupling. 

As was discussed below Eq.~\eqref{mh}, in the small $\mu$ region,  it is possible to raise the tree level mass of the lightest Higgs particle beyond the MSSM value by taking into account the additional contribution $(B_\mu v/f)^2 \sin^2 2\beta$ in Eq.~\eqref{mh}. 
While the MSSM tree level contribution, given by the first term in Eq.~\eqref{mh}, is bounded from above by $m_Z^2$, this additional contribution is bounded by the ratio  $(B_\mu v/f)^2$. This implies that it is possible to obtain a Higgs mass of $m_h =125 $ GeV already at tree level by taking $B_\mu/f$, as well as $\tan\beta,$ sufficiently close to one. However, even if the value of $B_\mu/f$ is such that it gives rise to a tree level Higgs mass which is below 125 GeV, the physical Higgs mass can still be compatible with such a value once moderate quantum corrections are taken into account.

In order to analyze  whether there exist regions of the parameter space where the limits in Fig.~\ref{Fig2} can be reached, we search for maxima of the quantities \sloppy\mbox{$\text{BR}(h\to \chi_{1}^{0}G)\times \text{BR}(\chi_{1}^{0}\to \g G)$} and  $h \chi^0_1 G$-coupling by scanning over the ranges $\mu\in[50,200]$ GeV, $m_1\in [50,200]$ GeV, $m_2\in [200,1000]$ GeV and $\tan\beta\in [1,4]$. We only take into account parameter regions which give rise to NLSP neutralino masses in the range $m_h/2 < m_{\chi_1^0} < m_h$ and a Higgs mass of \mbox{$m_h=125 $ GeV}. 

\begin{figure}[!t]
\begin{center}
\includegraphics[scale=0.36]{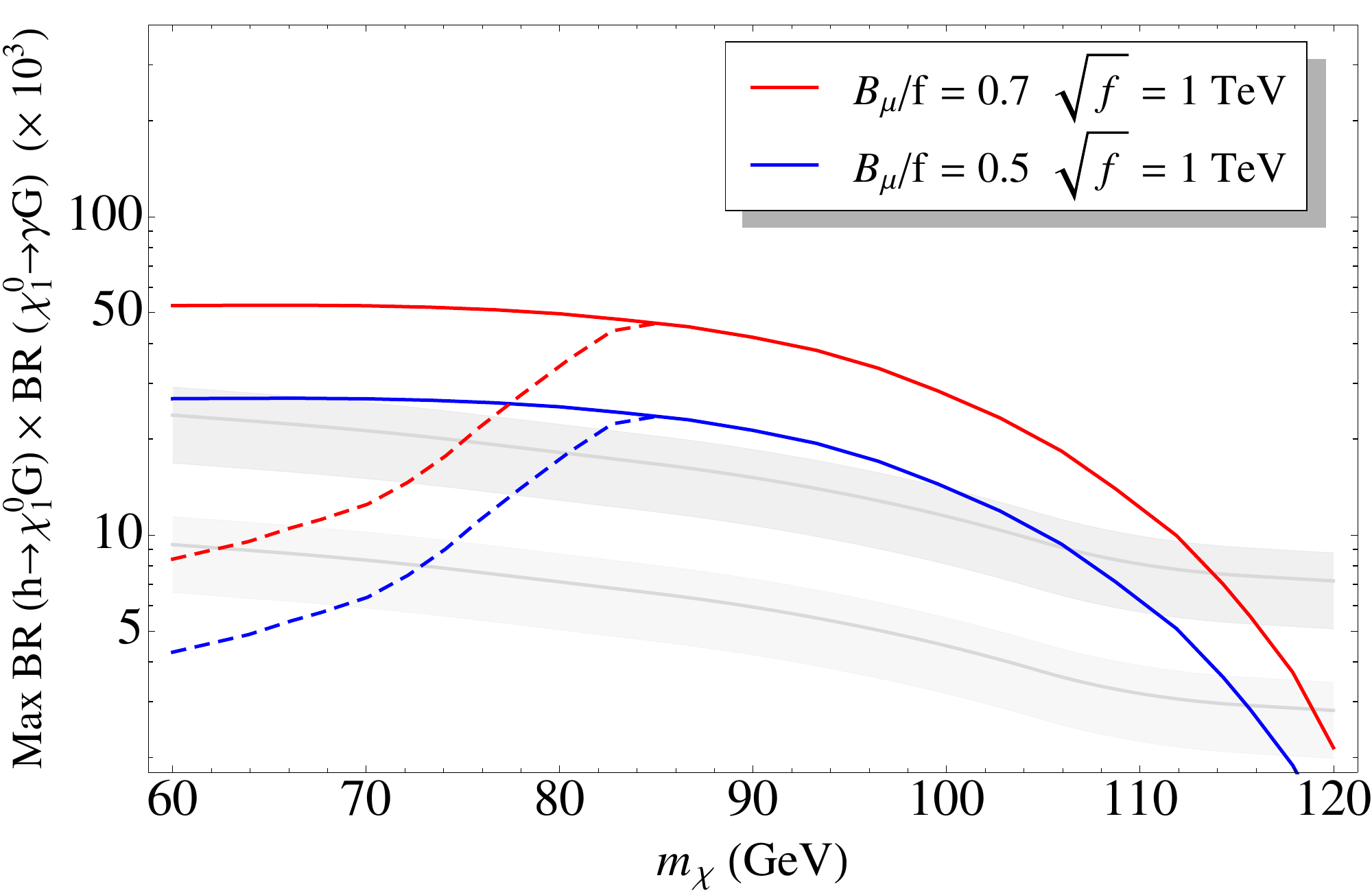}\hspace{2mm}
\includegraphics[scale=0.36]{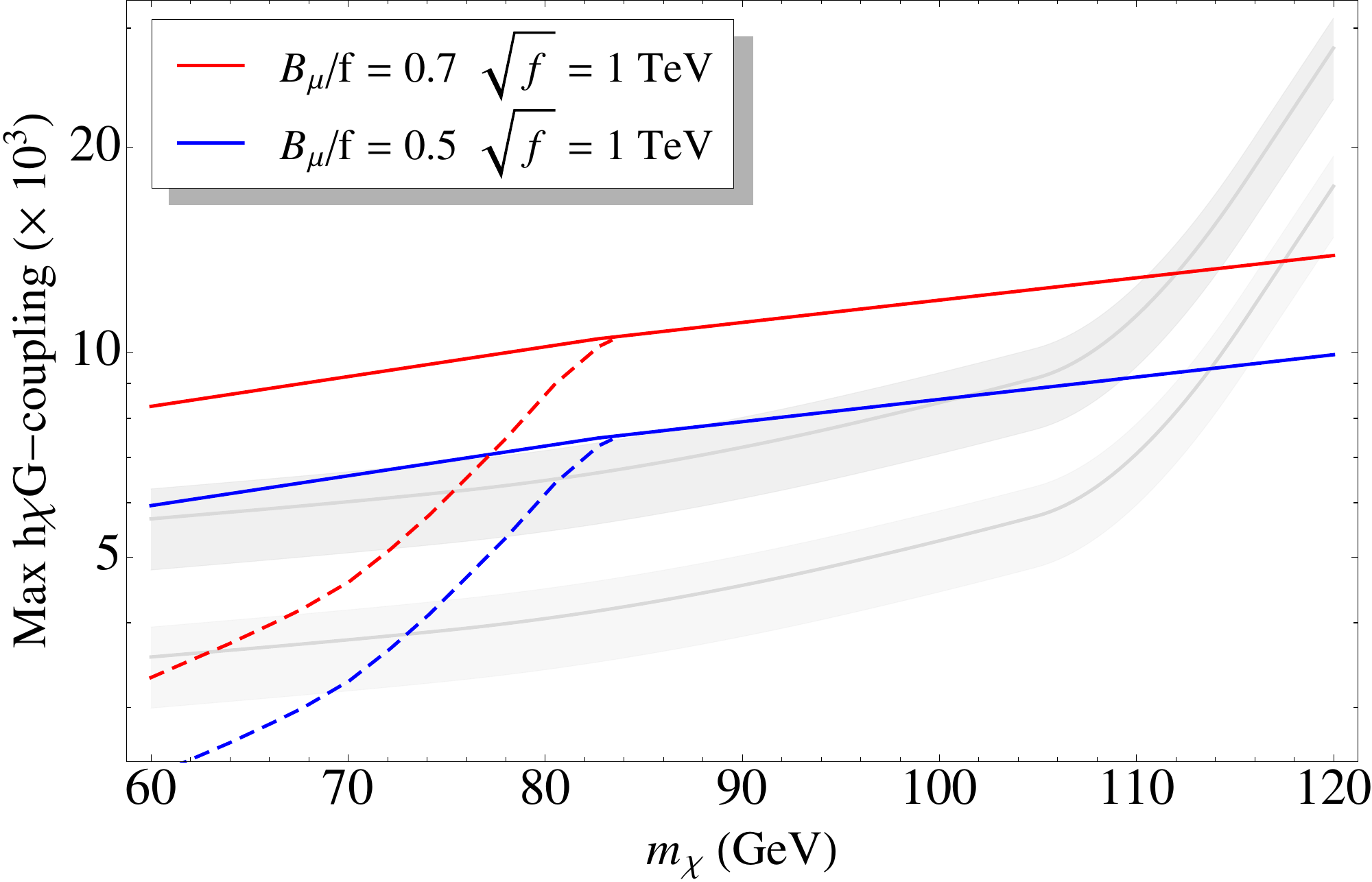}
\end{center}
\caption{ 
\small
The maximum values of $\text{BR}(h\to \chi_{1}^{0}G)\times \text{BR}(\chi_{1}^{0}\to \g G)$ (left panel) and the $h \chi^0_1 G$-coupling (right panel) that can be obtained in the model in Eq.~\eqref{hnG} . The discovery and exclusion limits in Fig.~\ref{Fig2} are shown as the gray lines and shaded bands. The solid red/blue curves corresponds to the chargino mass constraint $m_{\chi^{\pm}_i}>94$ GeV while the dashed red/blue curves corresponds to $m_{\chi^{\pm}_i}>103$ GeV. We have taken $\sqrt{f}=1$ TeV for all curves and $B_\mu/f=1$ and 0.7 for the red (upper) and blue (lower) ones, respectively.
}\label{Fig2p}
\end{figure}

The  $h \chi^0_1 G$-coupling of Eq.~\eqref{hnG} can be shown to increase for increasing values of  $B_\mu/f$ and therefore, the photon$\,+\,\MET\,$signal rate is favored by large values of  $B_\mu/f$. 
However it can be seen from Fig.~\ref{Fig2p} that already a value of $B_\mu/f=0.5$ is enough for discovery in a region of the parameter space. In this figure we have imposed the chargino masses of Eq.~\eqref{charginos} to be heavier than 94 GeV and 103 GeV \cite{Nakamura:2010dr}, corresponding to the solid and dashed curves, respectively.  In the lower half of the neutralino mass range we see that the maximum BR and coupling are reduced for the $m_{\chi^{\pm}_i}>103$ GeV curve with respect to the one with $m_{\chi^{\pm}_i}>94$ GeV, due to the difficulty in having the mass of the lightest neutralino significantly below the lightest chargino.
In addition, inspired by the commonly used value of the corresponding parameter $\lambda$ in the NMSSM, we also analyze the case where $B_\mu/f=0.7$. 
By using the fact that the $\text{BR}(h\to \chi_{1}^{0}G)\times \text{BR}(\chi_{1}^{0}\to \g G)$ scales as $1/f^2$ and the $h \chi^0_1 G$-coupling as $1/f$ it is easy to rescale the curves in Fig.~\ref{Fig2p} for other values of $f$. 

For both values of $B_{\mu}/f$ that we consider, the tree level Higgs mass is generically below 125 GeV. By taking into account the standard MSSM 1-loop correction with the stop mass in the range $m_{\tilde{t}}\in [200,500]$ GeV, with vanishing or moderate mixing, it is possible to obtain $m_h=125$ GeV (see Appendix \ref{appendix} for the relevant formulae). 

 In the lower part of the NLSP neutralino mass range, the mass of the second lightest neutralino $\chi^0_2$ can be below $m_h$, allowing for the decay $h\to\chi^0_2 G$. Since the dominant decay of $\chi^0_2$ is also into a photon and a goldstino, this scenario will actually increase the number of signal events. Let us also mention that the invisible Higgs decay $h\to GG$ is possible, but from Eqs.~\eqref{higgsvert} and \eqref{transf} we see that the $hGG$-coupling will be suppressed by at least a factor of $v\mu/f$ or $v^2/f$ with respect to the $h\chi^0_1 G$-coupling, implying that $\text{BR}(h\to GG)$ is very small with respect to $\text{BR}(h\to\chi^0_1 G)$.

\section{Conclusions}\l{Sec:Conclusion}

In this paper we studied the possibility of discovering an excess in the monophoton\,+\,$\MET$\,channel at the LHC and interpreting it as a signature for low scale SUSY breaking. In particular, in the region where the transverse momenta of the photon is $p_T^\gamma<m_h/2$, the 
dominant contribution to this channel can arise from a SM-like Higgs boson decay into an LSP goldstino and an NLSP neutralino, which promptly decays into another goldstino and a photon. In order for this decay to be kinematically allowed (but avoiding overlaps with the study of Higgs decays into two neutralinos in Refs.~\cite{Mason:1207862,Mason:2011uy}) we consider neutralino masses in the region $m_h/2<m_\chi<m_h$, where $m_h=125$ GeV. In order for the $\text{BR}(h\to \chi_{1}^{0}G)$ to be relevant, the neutralino must have a significant higgsino-component and the SUSY breaking scale $\sqrt{f}$ should be of the order of a few TeV. 
From our SM background analysis we concluded that the significance of the signal is optimized if the transverse momentum of the photon is in the region $m_Z/2<p_T^\gamma<m_h/2$.  Using suitable kinematic cuts, we estimated the prospects for a discovery or exclusion at the LHC with 8 TeV center of mass energy and 20 fb$^{-1}$ of integrated luminosity, as a function of $\text{BR}(h\to \chi_{1}^{0}G)$. We also discussed an effective model with manifest, but spontaneously broken, supersymmetry in which it is possible to obtain a SM-like Higgs boson of mass \mbox{$m_h=125$ GeV} and a value of $\text{BR}(h\to \chi_{1}^{0}G)$ sufficient for a discovery by the 2012 LHC run. 

In the effective model we considered, the usual MSSM soft terms are promoted to supersymmetric operators involving a dynamical goldstino supermultiplet. Since most of these operators are non-renormalizable, two natural questions arise. The first concerns the effect of higher dimensional operators not related to the MSSM soft parameters. As discussed in Section \ref{Sec:Model} these operators are not expected to significantly affect the relevant couplings while they can  affect the Higgs and neutralino mass spectrum. However, we expect that the introduction of additional higher dimensional operators (and the corresponding new parameters) could even improve our results. The second question concerns the possibility of embedding this effective model into an ultraviolet complete framework. Naively, one might attempt to embed it into the framework of gauge mediation, see Ref.~\cite{Giudice:1998fj} for a review. The problem is that the absence of tachyonic states in the messenger sector implies that $\sqrt{f}$ is in general at least one inverse loop factor above the soft mass parameters and therefore it seems difficult to obtain a viable standard gauge mediation scenario  with \mbox{$\sqrt{f}\lesssim 50$ TeV}. However, it might be possible to construct a viable microscopic model with a lower $\sqrt{f}$ by using elements of, for example, tree level gauge mediation \cite{Nardecchia:1207267} or general gauge mediation \cite{Meade:2008wd}. Another framework\footnote{Scenarios that allow for a gravitino mass corresponding to such a low $\sqrt{f}$ have also been discussed in the context of no-scale supergravity \cite{1984PhLB..147...99E}.} which involves models with $\sqrt{f}$ of the order of 1 TeV concerns five-dimensional warped spaces dual, via the AdS/CFT correspondence, to four dimensional models in which SUSY is broken by strong dynamics \cite{Gherghetta:2000br,Gherghetta:2000ez,Gherghetta:2011tn}. It would be interesting to explore the possibility of embedding the effective model considered in this paper, or a variation of it, into one of these frameworks and to study the phenomenological consequences of such a microscopic model.

\section*{Acknowledgments}
We thank E.~Dudas and G.~Ferretti for interesting conversations, R.~Franceschini for discussions on the final state considered in this paper, V.~Giangiobbe for discussions on  the relevant backgrounds, C.~Pena for computer support and D.~Tommasini for conversations concerning the radiative corrections to the Higgs production cross section. This work was supported by the Spanish MICINNÕs Juan de la Cierva  and Consolider-Ingenio 2010 programme under grants CPAN CSD2007- 00042, FPA2009-07908, FPA2010-17747, MultiDark CSD2009-00064, the Community of Madrid under grant HEPHACOS S2009/ESP-1473 and the European Union under the Marie Curie-ITN programme PITN-GA-2009-237920. The work of R.T. was partially supported by the Research Executive Agency (REA) of the European Union under the Grant Agreement number PITN-GA-2010-264564 (LHCPhenoNet).

\appendix

\section{Analytic formulae}\l{appendix}

The formula we have used in order to compute the mass of the lightest Higgs particle is given by,
\begin{equation}
\label{mhtot}
m_h^2 = m_{Z}^2 \cos^2 2\beta+\frac{v^2}{f^2}   \left(  2\mu^2-B_\mu \sin2\beta \right)^2  + 4 \,\delta\, v^2\, \sin^4 \beta~,
\end{equation}
where, in addition to the tree level contribution in Eq.~\eqref{mh}, we have here also included the leading MSSM 1-loop contribution parametrised by $\delta$. This corresponds to adding the term $V_{\text{1-loop}}=\delta|h_u|^4$ to the scalar potential in which,
 \begin{eqnarray}
\delta & = & \frac{3m_t^4}{16\pi^2v^4\sin^4 \beta}\Bigg[ \log \left( \frac{m_{\tilde{t}}^2}{m_t^2} \right) +\frac{X_t^2}{m_{\tilde{t}}^2}\left(1-\frac{X_t^2}{12m_{\tilde{t}}^2} \right)\Bigg]~,
\end{eqnarray}
where $X_t=A_t-\mu/\tan\beta$ and $m_{\tilde{t}}^2=m_{Q_{3}}m_{U^{c}_{3}}$. 

Once the dependence on  $\delta$ is included, the relevant vertices for Eqs.~\eqref{higgsvert}, \eqref{photonvert} and \eqref{Zvert} for the effective model in Section \ref{Sec:Model}, in the mass basis, are given by, 
\begin{align}
& \bry{lll}
\La_{h \chi G} &\supset& \dst\Bigg[\frac{1}{2f}\bigg( R_{(h,u)} \left(\sqrt{2} B_{\mu } N_{(1,d)}+g_1 v N_{(1,B)} (m_1 \sin \beta -\mu  \cos \beta )  \right.\vspace{2mm}\\
&&\dst \quad  \left.+ \sqrt{2} N_{(1,u)} \left(-B_{\mu } \cot \beta +2 v^2 \delta  \sin ^2\beta +\mu^2 \right)+g_2 v N_{(1,W)} (\mu  \cos \beta -m_2 \sin \beta )\right)  \vspace{2mm}\\
&&\dst \quad +R_{(h,d)} \left(\sqrt{2} N_{(1,d)} \left(\mu ^2-B_{\mu } \tan \beta \right)+g_1 v N_{(1,B)} (\mu  \sin \beta - m_1 \cos \beta ) \right. \vspace{2mm}\\
&&\dst \quad  \left. + \sqrt{2} B_{\mu } N_{(1,u)}+g_2 v N_{(1,W)} (m_2 \cos \beta -\mu  \sin \beta )\right) \bigg) \vspace{2mm}\\
&&\dst \quad +\frac{v^2 B_{\mu }^2} {\sqrt{2} f^3}\left(N_{(1,d)} R_{(h,d)}\sin ^2\beta  +N_{(1,u)} R_{(h,u)}\cos ^2\beta  \right)  \Bigg] h \chi G +\text{h.c.}~,\l{hnG}
\ery\\
& \bry{lll}
\mathcal{L}_{\chi\gamma G} & \supset & -
\dst \frac{1}{\sqrt{2}f}\Big[m_1 N_{(1,B)} \cos \theta_w +m_2 N_{(1,W)} \sin \theta_w  \Big] \chi_1^0 \sigma^{\mu\nu} F_{\mu\nu}  G +\text{h.c.}~,
\ery\\\nn\\
& \bry{lll}
\mathcal{L}_{\chi Z G} & \supset & 
\dst \frac{1}{\sqrt{2}f}\Big[m_1 N_{(1,B)} \sin \theta_w -m_2 N_{(1,W)} \cos \theta_w \Big] \chi_1^0 \sigma^{\mu\nu} Z_{\mu\nu}  G \vspace{2mm}\\
&& \dst +\frac{m_Z \mu}{\sqrt{2}f}\Big[  N_{(1,u)} \cos \beta  -N_{(1,d)} \sin \beta  \Big]\chi_1^0 \overline{\sigma}^\mu Z_\mu G
 +\text{h.c.}~,
\ery
\end{align}
where the matrix elements of the rotation matrix $R$ for the neutral scalars are given by,
\begin{eqnarray}
\label{R}
&&R_{(h,d)}= \frac{\cos \beta }{B_{\mu } } \Big(B_{\mu }-2 \sin ^3\beta  \cos \beta  \left(\cos 2 \beta  \left(m_Z^2+v^2 \delta \right)-v^2 \delta \Big)\right)
\nonumber \\
&& \qquad \qquad +\frac{v^2 B_{\mu }}{f^2} \sin ^3\beta  \cos \beta  (\cos \beta +\cos 3 \beta ) ~,
\nonumber \\
&& R_{(h,u)}=  \frac{\sin \beta }{B_{\mu } }\Big(B_{\mu }+2 \sin \beta  \cos ^3\beta  \left(\cos 2 \beta  \left(m_Z^2+v^2 \delta \right)-v^2 \delta \right)\Big)
\nonumber \\
&& \qquad \qquad +  \frac{v^2 B_{\mu }}{f^2}  \sin \beta  \cos ^3\beta   (\sin \beta -\sin 3 \beta )
\end{eqnarray}
and the analytic formulae for the matrix elements of the rotation matrix $N$ for the neutralinos are given in Ref.~\cite{ElKheishen:1991gx}.

\def\bstname{fdp}

%
 \bibliographystyle{jhep}
\bibliography{paper}{}

\end{document}